\documentclass{aa}
\usepackage{latexsym}
\usepackage{epsfig}
\usepackage{graphicx}
\newcommand{\radm}{rad~m$^{-2}$} 
\newcommand{\pl}{\parallel}
\newcommand{\dg}{\degr}
\newcommand{\srm}{\mbox{$\sigma_{\scriptscriptstyle\rm RM}$}}
\renewcommand{\ga}{\,\hbox{\rlap{\hbox{\lower4pt\hbox{$\sim$}}}
                  \lower-1pt\hbox{\hspace{-4pt}$>$}}\,}
\begin{document}

\title{Multi-frequency polarimetry of the Galactic radio background
        around 350~MHz: I. A region in Auriga around l = 161\dg, b =
        16\dg}

   \author{M. Haverkorn\inst{1}
          \and
          P. Katgert\inst{2}
	  \and
	  A. G. de Bruyn\inst{3,4}
          }
   \offprints{M. Haverkorn}
   \institute{Leiden Observatory, P.O.Box 9513, 2300 RA Leiden, the
              Netherlands\\
	      (Current address: Harvard-Smithsonian Center for
	      Astrophysics, 60 Garden Street MS-67, Cambridge MA
	      02138, USA)\\
              \email{mhaverkorn@cfa.harvard.edu} 
         \and Leiden Observatory, P.O.Box 9513, 2300 RA Leiden, the
              Netherlands\\ 
              \email{katgert@strw.leidenuniv.nl} 
         \and ASTRON, P.O.Box 2, 7990 AA Dwingeloo, the Netherlands\\ 
              \email{ger@astron.nl}
         \and Kapteyn Institute, P.O.Box 800, 9700 AV Groningen,
              the Netherlands }
\date{Received, accepted}
\titlerunning{Multi-frequency polarimetry around 350~MHz I}

\abstract{
With the Westerbork Synthesis Radio Telescope (WSRT), multi-frequency
polarimetric images were taken of the diffuse radio synchrotron
background in a $\sim$~5\dg$\times$7\dg\ region centered on $(l,b) =
(161\dg,16\dg)$ in the constellation of Auriga.  The observations were
done simultaneously in 5 frequency bands, from 341~MHz to 375~MHz, and
have a resolution of 
$\sim$~5.0\arcmin$\times$5.0\arcmin~cosec~$\delta$.  The polarized
intensity $P$\/ and polarization angle $\phi$ show ubiquitous
structure on arcminute and degree scales, with polarized brightness 
temperatures up to about 13~K. On the other hand, no structure at all
is observed in total intensity $I$ to an r.m.s.\ limit of 1.3~K, 
indicating that the structure in the polarized radiation must be due to
Faraday rotation and depolarization mostly in the warm component of
the nearby Galactic interstellar medium (ISM). Different
depolarization processes create structure in polarized intensity
$P$. Beam depolarization creates ``depolarization canals'' of one beam
wide, while depth depolarization is thought to be responsible for
creating most of the structure on scales larger than a beam width.
Rotation measures ($RM$) can be reliably determined, and are in the
range $-17 \la RM \la 10$~\radm\ with a non-zero average $RM_0 \approx
-3.4$~\radm. The distribution of $RM$s on the sky shows both abrupt
changes on the scales of the beam and a gradient in the direction of
positive Galactic longitude of $\sim$~1~\radm\ per degree. The
gradient and average $RM$ are consistent with a regular magnetic field
of $\sim 1~\mu$G which has a pitch angle of $p = -14$\dg.  There are
13 extragalactic sources in the field for which $RM$s could be
derived, and those have $|RM| \la 13$~\radm, with an estimated
intrinsic source contribution of $\sim$~3.6~\radm.  The $RM$s of the
extragalactic sources show a gradient that is about 3 times larger
than the gradient in the $RM$s of the diffuse emission and that is
approximately in Galactic latitude. This difference is ascribed to a
vastly different effective length of the line of sight. The $RM$s of
the extragalactic sources also show a sign reversal which implies a
reversal of the magnetic field across the region on scales larger than
about ten degrees. The observations are interpreted in terms of a
simple single-cell-size model of the warm ISM which contains gas and
magnetic fields, with a polarized background. The observations are
best fitted with a cell size of 10 to 20 pc and a ratio of random to
regular magnetic fields $B_{ran}/B_{reg} \approx 0.7\pm0.5$. The
polarization horizon, beyond which most diffuse polarized emission is
depolarized, is estimated to be at a distance of about 600~pc. 

}

\maketitle

\section{Introduction}
\label{s4:intro}

The warm ionized gas and magnetic field in the Galactic disk
Faraday-rotate and depolarize the linearly polarized component of the
Galactic synchrotron emission.  Structure of the warm ionized
Interstellar Medium (ISM) is imprinted in the polarization angle and
polarized intensity by Faraday rotation and depolarization. Therefore,
the structure in the polarized diffuse radio background provides
unique information about the structure in the magneto-ionized
component of the ISM. Using interferometers we can probe arcminute
scales, which correspond to linear scales from a fraction of a parsec
to several tens of parsecs. Multi-frequency radio polarization
observations of the Faraday-rotated synchrotron emission enable
determination of the rotation measure ($RM$) distribution along many
contiguous lines of sight. Therefore, this is a valuable method to
estimate the non-uniform component of the Galactic magnetic field,
weighted with electron density.

Although the magnetic field is only one of the factors that shape the
very complex multi-component ISM, it is generally thought to play a
major r\^ole in the energy balance of the ISM and in setting up and
maintaining the turbulent character of the medium.  The non-uniform
component of the magnetic field may influence heating of the ISM
(Minter \& Balser 1997), and provide global support of molecular
clouds (e.g.\ V\'azquez-Semadeni et al.\ 2000, Heitsch et al.\
2001). It also plays an important r\^ole in star formation processes
(e.g. Shu 1985, Beck et al.\ 1996, Ferri\`ere 2001). Furthermore,
observations of the detailed structure of the magnetic field can
provide constraints for Galactic dynamo models (Han et al.\ 1997).

The strength and structure of the magnetic field in the warm ionized
medium in the Galaxy can also be obtained from Faraday rotation
measurements of pulsars, or polarized extragalactic sources.

Among these, radio observations of pulsars take a special place
because for them one can measure the $RM$ and the dispersion measure
($DM$) along the same line of sight. The ratio of the two immediately
yields the averaged component of the magnetic field along a particular
line of sight (see e.g.\ Lyne \& Smith 1989, Han et al.\ 1999).
These studies, and analysis of the distribution of polarized
extragalactic sources (e.g. Simard-Normandin \& Kronberg 1980, Sofue
\& Fujimoto 1983), suggest that the regular Galactic magnetic field
is directed along the spiral arms, with a pitch angle of $p \approx 5
- 15$\dg. There is evidence for reversals of the uniform magnetic
field inside the solar circle (e.g.\ Simard-Normandin \& Kronberg
1979, Rand \& Lyne 1994), while there is still debate on reversals
outside the solar circle (e.g. Vall\'ee 1983, Brown \& Taylor 2001,
Han et al.\ 1999).  Rand \& Kulkarni (1989) used pulsar $RM$s to derive
a value for the strength of the regular component of the magnetic
field of \mbox{$B_{reg} = 1.3 \pm 0.2~\mu$G}.  This result assumes a
circularly symmetric large-scale Galactic magnetic field. The
residuals with respect to the best-fit large-scale model were
interpreted as due to a random magnetic field of 5~$\mu$G. This number
was derived by assuming a scale length of the random field, modeled as
a cell size, of 55~pc.  Ohno \& Shibata (1993) confirmed this result
for the random magnetic field component for all cell sizes in the
range of 10~-~100~pc. They used 182 pulsars in pairs and therefore did
not have to make any assumptions about a large-scale Galactic magnetic
field.

Using extragalactic sources, most of which are double-lobed with lobe
separations between 30\arcmin\arcmin\ and 200\arcmin\arcmin, Clegg et
al.\ (1992) found substantial fluctuations in $RM$ on linear scales of
$\sim$~0.1~-~10~pc, which they could explain by electron density
fluctuations alone. However, Minter \& Spangler (1996) observed $RM$
fluctuations in extragalactic source components that cannot be
explained by only electron density fluctuations; they need an
additional turbulent magnetic field of $\sim 1 \,\mu$G to fit the
observations. Observations of polarization of starlight (Jones et
al. 1992) give much larger estimates of the cell size, of up to a kpc.

Using pulsars and extragalactic radio sources for the determination of
the $RM$ of the Galactic ISM is clearly not ideal, because they only
provide information in particular directions which sample the ISM very
sparsely. On the contrary, the {\it diffuse} Galactic radio background
provides essentially complete filling over large solid
angles. Therefore, rotation measure maps of the diffuse emission can
be produced that give information on the electron-density-weighted
magnetic field over a large range of scales. The distribution of
polarized intensity, when interpreted as mostly being due to
depolarization, can yield estimates of several properties of the warm
ISM such as correlation length, ratio of random over regular magnetic
field and the distance out to which diffuse polarization can be
observed. Early $RM$ maps of the diffuse synchrotron emission in the
Galaxy were constructed by Bingham \& Shakeshaft (1967) and Brouw \&
Spoelstra (1976) who confirmed a Galactic magnetic field in the
Galactic plane. Junkes et al.\ (1987) presented a polarization survey
of the Galactic plane at $4.9\dg < l < 76\dg$ and $|b| < 1.5\dg$
showing small-scale structure in diffuse polarization. The existence
of polarization filaments at intermediate latitudes, without
correlated structure in total intensity $I$, was discovered by
Wieringa et al.\ (1993) at 325~MHz. Polarization surveys have been
performed at frequencies from 1.4~GHz to 2.695~GHz (Duncan et al.\
1997, Duncan et al.\ 1999, Uyan\i ker et al.\ 1999, Landecker et al.\
2001, Gaensler et al.\ 2001), mostly in the Galactic plane.

In this paper, we discuss the results of multi-frequency observations
at low frequencies around 350 MHz of a field in the constellation
Auriga, in the second Galactic quadrant ($l = 161^{\circ}$, $b =
16^{\circ}$). Due to the low frequencies, we probe low rotation
measures that are predominant at intermediate and high latitudes. This
gives the opportunity to study the high-latitude $RM$ without concrete
objects such as HII regions or supernova remnants in the line of
sight, and estimate structure in the ISM above the thin stellar disk.

Section~\ref{s4:obs} contains details of the multi-frequency
polarization observations. In Sect.~\ref{s4:res}, we analyze the
observations, in particular the small-scale structure in polarized
intensity and polarization angle. Faraday rotation is discussed in
Sect.~\ref{s4:rm}, where we also present the map of rotation
measure.  In Sect.~\ref{s4:depol}, depolarization mechanisms are
described that cause the structure in $P$, and the constraints that
the observations provide for the parameters that describe the warm
ISM. In Sect.~\ref{s4:sources}, the polarization properties of 13
polarized extragalactic point sources found in the Auriga region are
discussed. In Sect.~\ref{s4:disc}, we discuss the information that
our data provides on the strength and structure of the Galactic
magnetic field.  Finally, our conclusions are stated in
Sect.~\ref{s4:conc}.

\section{The observations}
\label{s4:obs}

We used the Westerbork Synthesis Radio Telescope (WSRT) for
multi-frequency polarimetric observations of the Galactic radio
background in a field in the constellation of Auriga centered on
($\alpha$, $\delta$) (B1950)~=~$(6^h20^m, 52^{\circ}30^m)$ ($l =
161^{\circ}$, $b = 16^{\circ}$).  This field was observed in 8
frequency bands between 325 and 390~MHz simultaneously, each with a
band width of 5~MHz. Due to radio interference and hardware problems
only data in 5 of the 8 bands, viz.\ those centered
on 341, 349, 355, 360, and 375~MHz, could be used.

The region in Auriga was observed in six 12hr periods, which resulted
in a baseline increment of 12m.  The shortest baseline obtained is
36m, and the longest is 2700m, which gives a maximum resolution of
$\sim 1$\arcmin. A taper was applied to the $(u,v)$-data to increase
the signal-to-noise ratio, so as to obtain a resolution of
$5.0\arcmin\times5.0\arcmin$~cosec~$\delta =
5.0\arcmin\times6.3\arcmin$ in all 5 frequency bands.  As an
interferometer has a finite shortest baseline, it is insensitive to
large-scale structure. The shortest spacing of 36m in the WSRT
constitutes effectively a high-pass filter for all scales above
approximately a degree. 

The Auriga region was selected from diffuse polarization maps that
were produced as a by-product of the Westerbork Northern Sky Survey
(WENSS, Rengelink et al. 1997), which is a single-frequency radio
survey at 325~MHz. The WENSS diffuse polarization maps contain several
regions of high polarization which show conspicuous small-scale
structure in polarized intensity and polarization angle. We reobserved
two of those regions at multiple frequencies and with higher
sensitivity to obtain rotation measure information. The first region
in Auriga is described in this paper, and the other one in a
forthcoming paper (Haverkorn et al.\ 2003a). 

The data reduction process is described in detail in Haverkorn (2002),
and we only give a brief summary here.  The observations were reduced
using the {\sc newstar} data reduction package. Polarized and
unpolarized standard calibrator sources were used, where the absolute
flux scale at 325~MHz is based on a value of 26.93~Jy for 3C286 (Baars
et al.\ 1977). From this value the flux scales of the other calibrator
sources 3C48, 3C147, 3C345, and 3C303 were derived.

As the area to be mapped is larger than the primary beam of the WSRT,
the mosaicking technique was used (Rengelink et al.\ 1997). In this
observing mode, the array cycles through a preselected set of pointing
positions a number of times during the 12 hour observation period,
which reduces the instrumental polarization in between pointing
centers to below 1\%.  For the present region we used 5$\times$7
pointing positions separated by 1.25\dg. We only discuss the central
$\sim 7\dg\times$9\dg\ of the observed field of view, as the edges of
the mosaic exhibit larger noise due to primary beam attenuation and
increased instrumental polarization.  Maps were constructed of Stokes
parameters $I$, $Q$, $U$, and $V$. From Stokes $Q$ and $U$, polarized
intensity $P=\sqrt{Q^2 + U^2}$ and polarization angle $\phi =
0.5\arctan(U/Q)$\footnote{Note that this notation, although much used,
is actually incorrect for $Q<0$. The correct notation is $\phi =
0.5\,\mbox{angle}(Q,U)$, or $\phi = 0.5\arctan(U/Q)$ for $Q>0$ and $\phi
= 0.5\arctan(U/Q) + 0.5*\pi$ for $Q<0$.} were derived.

\begin{table}
  \begin{center}
    \begin{tabular}{|l|ccc|}
      \hline
  Central position & \multicolumn{3}{l|}{(l,b) = (161\dg, 16\dg)} \\
  Size             & \multicolumn{3}{l|}{$\sim$~7\dg$\times$9\dg}    \\  
  Pointings        & \multicolumn{3}{l|}{5$\times$7}                  \\  
  Frequencies      & \multicolumn{3}{l|}{341, 349, 355, 360, 375 MHz} \\ 
  Resolution       &  
  \multicolumn{3}{l|}{5.0\arcmin$\times$5.0\arcmin\ cosec $\delta$ 
                    = 5.0\arcmin$\times$6.3\arcmin}                 \\ 
  Noise            & \multicolumn{3}{l|}{$\sim$~4~mJy/beam (0.5~K)}   \\ 
  Conversion Jy - K& \multicolumn{3}{l|}{1 mJy/beam = 0.127 K (at 350~MHz)}\\
  \hline\hline
        Spacings & Date     & Start time & End time\\
                 &          &       (UT) &     (UT)\\
  \hline
        36m      & 96/02/19 & 13:36      & 01:54 \\ 
        48m      & 95/12/24 & 17:28      & 05:41\\
        60m      & 95/12/31 & 17:13      & 05:31\\
        72m      & 96/01/07 & 16:37      & 04:55 \\
        84m      & 96/01/30 & 14:59      & 03:12\\
             96m & 96/01/09 & 16:33      & 04:46\\  
  \hline
    \end{tabular}
    \caption{Observational details of the Auriga field}
    \label{t4:data}
  \end{center}
\end{table}

\begin{figure}
\centering
\psfig{figure=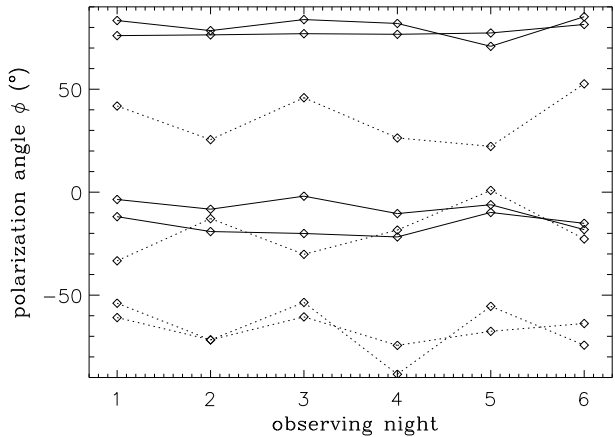,width=.5\textwidth}
\caption{Polarization angle variation over six 12hr observations, for
8 extragalactic point sources. Four sources denoted by the solid lines
are relatively strong ($\sigma$ = 2 - 3.5 for one 12hr measurement),
the dotted lines are sources with $\sigma$ = 1.2 - 1.5. Little angle
variation over nights points to constant ionospheric Faraday
rotation.}
\label{f4:frion}
\end{figure}

To avoid spurious solar emission coming in through polarized side
lobes, all observations were done at night, as shown in
Table.~\ref{t4:data}. In addition, the observing period was close to
solar minimum and in winter, minimizing ionospheric Faraday
rotation. The high polarized brightness that we observe (see
Sect.~\ref{s4:p}) indicates that ionospheric Faraday rotation does
not vary much during each of the 12hr observations, because if it did,
hardly any polarized intensity would have resulted.  The differences
between the average ionospheric Faraday rotation in the six 12hr
observations were determined from several point sources (in the $Q$-
and $U$-maps at $\sim 1$\arcmin\ resolution).  The result is shown in
Fig.~\ref{f4:frion} for 4 strong sources and 4 weaker sources. The
strong sources have a signal-to-noise $\sigma$ = 2 - 3.5 for an
individual 12hr observation, which corresponds to an uncertainty in
polarization angle $\sigma_\phi \approx 10$\dg. The signal-to-noise of
the weaker sources is $\sigma$ = 1.2 - 1.5 ($\sigma_\phi \approx
22$\dg).  From Fig.~\ref{f4:frion} we conclude that differential
Faraday rotation between the six 12hr observations is negligible to
within the uncertainties with which it can be measured from the point
sources.

We can estimate the $RM$ contribution of the ionosphere from the earth
magnetic field and the electron density in the ionosphere. The
contribution of the ionospheric Faraday rotation is estimated as
follows. The total electron content (TEC) in the ionosphere above
Westerbork at night, in a solar minimum, and in winter is minimal:
TEC~$\approx 2.2 \; 10^{16}$~electrons~m$^{-2}$ (Campbell, private
communication).
The Auriga field at declination $\delta = 52.5$\dg\ is almost in the
zenith at Westerbork at hour angle zero. If we assume a vertical
component of the earth magnetic field of 0.5~G towards us, and a path
length through the ionosphere of 300~km, the $RM$ caused by the
ionosphere is 0.3~\radm\ at hour angle zero. At larger hour angles,
this is even less. So we expect the rotation measure values given in
this paper not to be affected by ionospheric Faraday rotation by more
than 0.5~\radm. 

\subsection{Missing large scale structure}
\label{ss4:off}

An interferometer is increasingly insensitive to structure on larger
angular scales due to missing small spacings. This means for the WSRT
that scales above approximately a degree cannot be detected. The $Q$
and $U$ maps are constructed so that in each mosaic pointing, the map
integral of $Q$ and $U$ are zero. This leads to missing large-scale
components in $Q$ and $U$, and therefore erroneous determinations of
$P$, $\phi$ and $RM$. However, the distribution of $RM$ is broad
enough within one pointing. Therefore, the variation in polarization
angle is so large that the average $Q$ and $U$ are close to zero, so
that missing large-scale components are negligible
(Sect.~\ref{s4:depol}).

\section{Total intensity and linear polarization maps}
\label{s4:res}

\subsection{Total intensity}
\label{s4:i}

In the left-hand panel of Fig.~\ref{f4:ipi} we show the map of total
intensity $I$\/ at 349~MHz with a resolution of
5.0\arcmin$\times$6.3\arcmin. Point sources were removed down to the
confusion limit of 5~mJy/beam. $I$\/ has a Gaussian
distribution around zero with width $\sigma_I = 1.3$~K (while the
noise, computed from $V$ maps, is $\sigma_V = 0.5$~K), and is
distributed around zero due to the insensitivity of the interferometer
to scales $\ga 1\dg$. From the continuum
single-dish survey at 408~MHz (Haslam et al.\ 1981, 1982), the
$I$-background at 408~MHz in this region of the sky is estimated to be
$\sim$~33~K with a temperature uncertainty of $\sim$~10\% and
including the 2.7~K contribution of the cosmic microwave background.
Assuming a temperature spectral index of $-2.7$, the backgrounds at
the lowest (341~MHz) and the highest (375~MHz) frequency of
observation are $\sim$~49~K and $\sim$~38~K, respectively. Of these,
approximately 25\% is due to sources, as estimated from source counts
(Bridle et al.\ 1972), and assuming the spectral index of the Galactic
background and the extragalactic sources to be identical. We thus
estimate the temperatures of the diffuse Galactic background at 341 and
375~MHz to be 37 and 29~K, respectively.

\subsection{Polarized intensity}
\label{s4:p}

In the right-hand panel of Fig. 2 we show the map of polarized
intensity at 349 MHz, with a resolution of
5.0\arcmin$\times$6.3\arcmin. The noise in this map is $\sim$~0.5~K,
and the polarized brightness temperature goes up to $\sim$~13~K, with
an average value of $\sim$~2.3~K.

\begin{figure*}
\begin{center}
\hbox{\psfig{figure=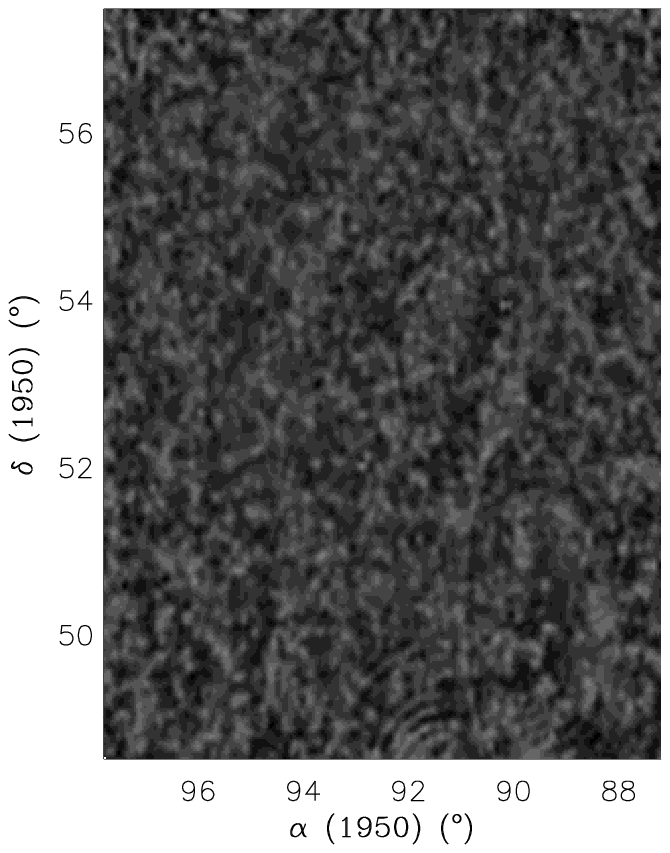 ,width=0.5\textwidth}
      \psfig{figure=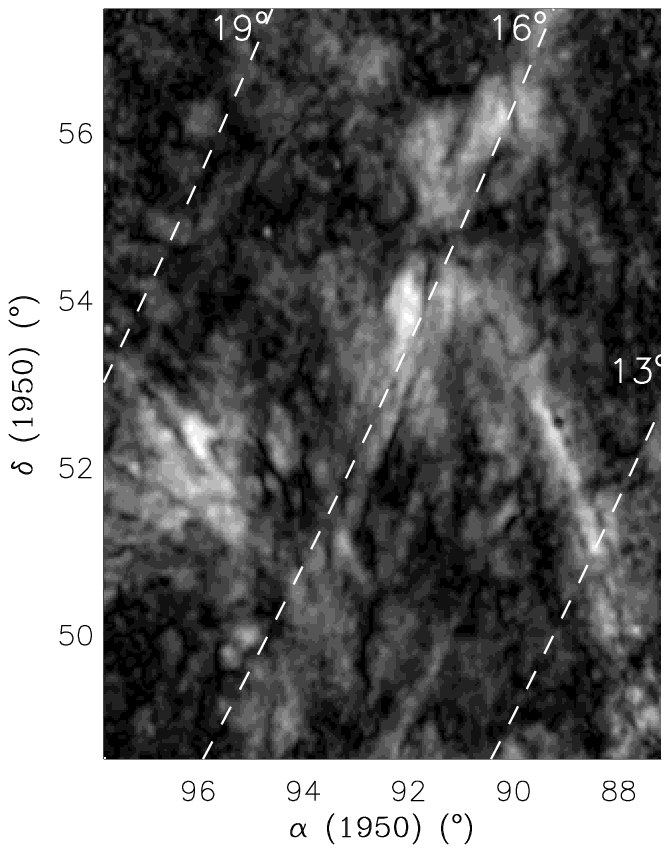,width=0.5\textwidth}}
\caption{Map of the total intensity $I$\/ (left) and polarized
intensity $P$\/ (right) at 349~MHz, on the same brightness scale. The
maps are smoothed to a resolution of $5.0^{\prime}\times
6.3^{\prime}$. White denotes high intensity, up to a maximum of
$\sim$~13~K. Superimposed in the right map are lines of constant
Galactic latitude, at latitudes of 13\dg, 16\dg\ and 19\dg.}
\label{f4:ipi}
\end{center}
\end{figure*}

The map shows cloudy structure of a degree to a few degrees in extent,
with linear features that are sometimes parallel to the Galactic
plane.  In addition, a pattern of black narrow wiggly canals is
visible (see e.g.\ the canal around
($\alpha$,~$\delta$)~=~(92.7$^{\circ}$, 49$^{\circ}$~-~51$^{\circ}$)).
These canals are all one synthesized beam wide and are most
likely due to beam depolarization. They separate regions of fairly
constant polarization angle between which the difference in
polarization angle is approximately 90\dg\ ($\pm\, n\, 180$\dg,
$n=1,2,3\ldots$) (Haverkorn et al. 2000). A change of 90$^{\circ}$ (or
270$^{\circ}$, 450$^{\circ}$ etc.) in polarization angle {\it within
one beam} cancels the polarized intensity in the beam; therefore these
canals appear black in Fig.~\ref{f4:ipi}. Hence, the canals are not
physical structures, but the reflection of specific features in the
distribution of polarization angle. Such canals have also been
detected by others at higher frequencies (e.g. Uyan\i ker et al. 1999,
Gaensler et al.\ 2001).  It must be appreciated that the canals
represent an extreme case of beam depolarization. Beam depolarization
is much more wide-spread if the polarization angle varies on the scale
of the beam or on smaller scales, but in general its effect is (much)
less than in the canals. Therefore, with the exception of the canals,
beam depolarization does not leave easily visible traces in the
polarized intensity distribution.

An alternative explanation for depolarized regions accompanied
by a change in polarization angle of 90\dg is so-called depth
depolarization or differential Faraday rotation. If the magnetic field
in the medium is uniform, depolarization along the line of sight
causes depolarized regions at certain $RM$ values (Sokoloff et
al.\ 1998). However, in this scenario, it is hard to explain why the
canals are all exactly one beam wide. Furthermore, the canals should shift in
position with frequency, which is not observed. For an extended
discussion on the possible causes of polarization canals, see
Haverkorn et al.\ 2003b.

There is no obvious structure in Stokes I, and if there is any, it
does not appear to be correlated with the structure in P. This is not
typical for the Auriga field, but it is true in all fields observed so
far with the WSRT at $\sim$~350~MHz (Katgert \& de Bruyn 1999,
Haverkorn et al.\ 2003a, Schnitzeler et al.\ in prep).

We have estimated the amount of correlation between total and
polarized intensity by deriving the correlation coefficient $C$, where
$C(f,g)$ of the observables $f(x,y)$ and $g(x,y)$ is defined as
\begin{equation}
  C(f,g) = \frac{\sum_{i,j} (f_{ij}-\overline{f})\;(g_{ij}-\overline{g})}
                {\sqrt{\sum_{i,j} (f_{ij} - \overline{f})^2 \;
       		       \sum_{i,j} (g_{ij} - \overline{g})^2}}
\end{equation}
where $\overline{f}$ is the average value of $f(x,y)$, $f_{ij} =
f(x_i,y_j)$, summation is over $i = 1, N_x$ and $j = 1, N_y$, and
$N_x$ and $N_y$ is the number of data points in the $x$- and
$y$-directions, respectively. The correlation coefficients between
$P$\/ and $I$, and between $P$ and between $I$ at different
frequencies are given in Table~\ref{t4:c}. Both the observed
correlation coefficients at the maximum resolution of about 1\arcmin,
as well as those of the 5\arcmin\ resolution data are given.

\begin{table}
\begin{center}
  \caption{Correlation coefficients for correlations between $P$\/ and
           $I$, between $I$\/ in different frequency bands and between
           $P$\/ in different frequency bands, for tapered data and full
           resolution data.}
  \label{t4:c} 
  \begin{tabular}{c|c|rrrrr} 
    freq band k & res &  $C(I_k,P_k)$ & $C(P_{341}, P_k)$ 
                             & $C(I_{341},I_k)$ \\
    (MHz)  & (\,\arcmin\,)     & & & \\
    \hline
    341 & 5 & --0.008  & 1.00 & 1.00 \\
    349 & 5 & --0.008  & 0.67 & 0.45 \\
    355 & 5 & --0.011  & 0.65 & 0.34 \\
    360 & 5 &   0.002  & 0.58 & 0.31 \\
    375 & 5 &   0.015  & 0.53 & 0.24 \\ \hline
    341 & 1 &   0.007  & 1.00 & 1.00 \\
    349 & 1 &   0.004  & 0.60 & 0.08 \\
    355 & 1 &   0.0004 & 0.60 & 0.21 \\
    360 & 1 &   0.0002 & 0.53 & 0.15 \\
    375 & 1 &   0.005  & 0.50 & 0.16 \\
  \end{tabular}
\end{center}
\end{table}

The high correlations between $I$\/ in different frequency bands are
expected if the visible structure in $I$\/ is mostly due to faint
sources. The $C(I,I) < 1$ is due to uncorrelated noise. However, the
correlation between $P$\/ and $I$\/ is very low, so the structure in
polarization is not due to small-scale variations in synchrotron
emission. Instead, the variations in polarized intensity and
polarization angle must be due to a combination of two processes:
Faraday rotation and depolarization. We will return to these in
Sects.~\ref{s4:rm} and~\ref{s4:depol}.

\begin{figure*}
\centerline{
\hbox{\psfig{figure=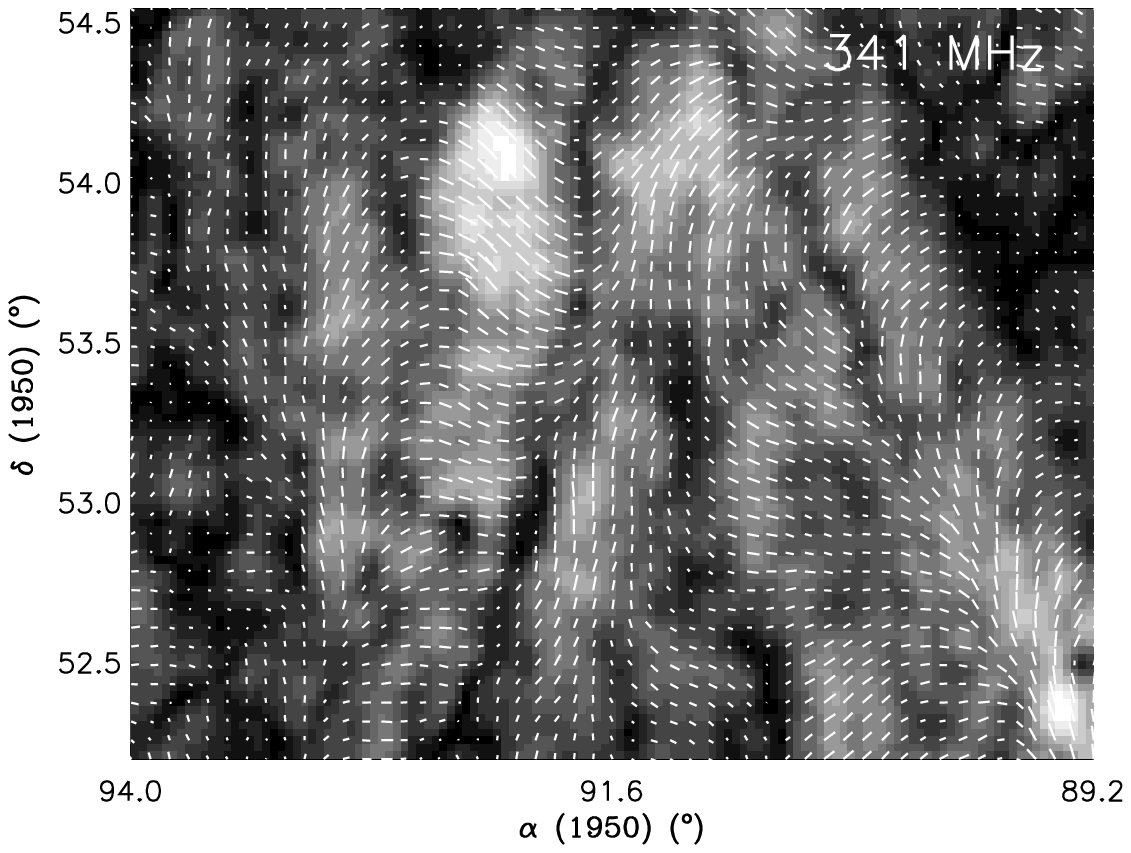,width=0.45\textwidth}
      \psfig{figure=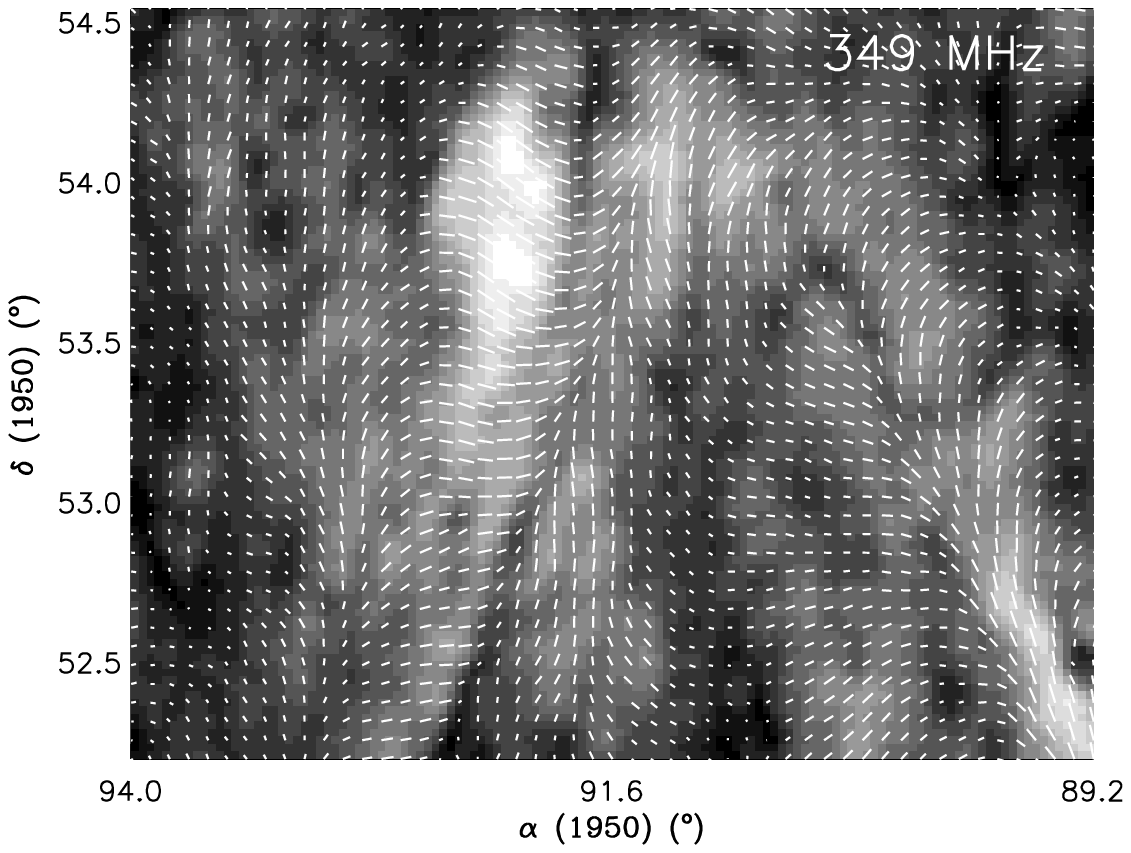,width=0.45\textwidth}}}
\centerline{
\hbox{\psfig{figure=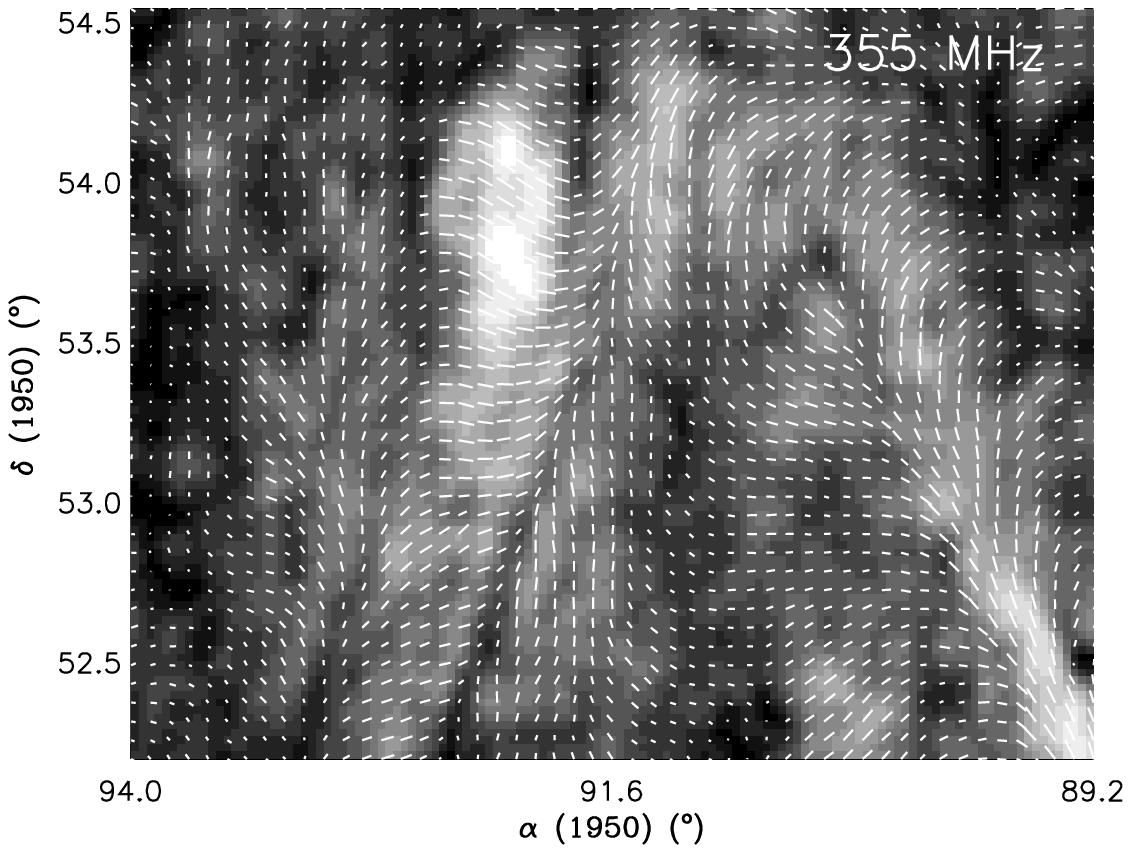,width=0.45\textwidth}
      \psfig{figure=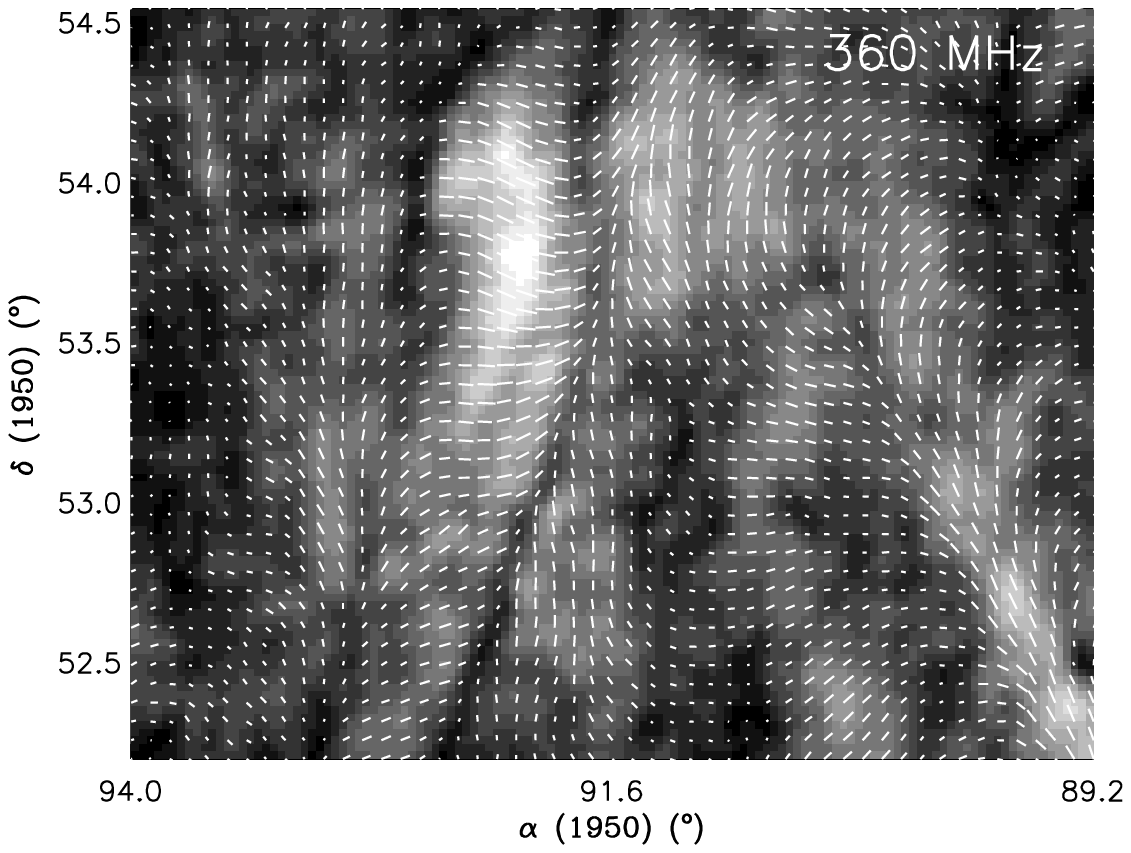,width=0.45\textwidth}}}
\centerline{
\hbox{\psfig{figure=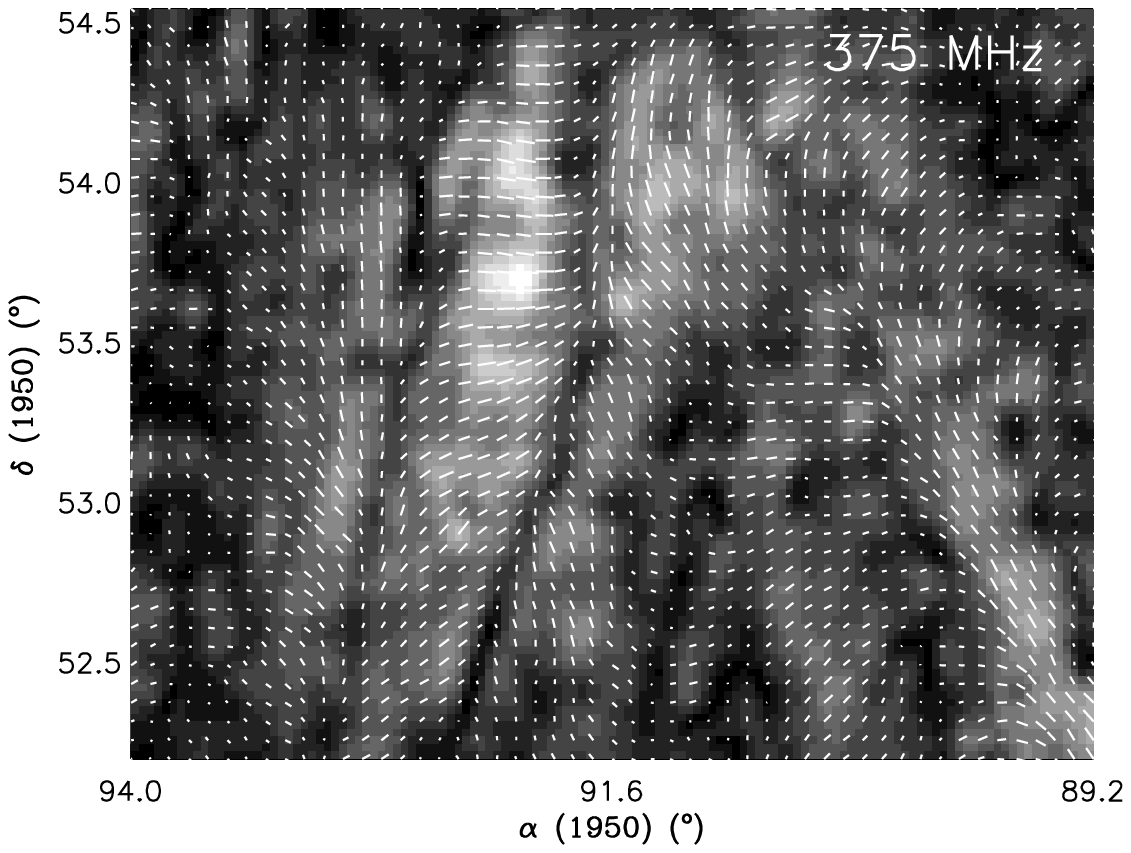,width=0.45\textwidth}
      \parbox[b]{0.45\textwidth}{\mbox{}}}}
\caption{Five maps of a part of the Auriga region at 
         frequencies 341, 349, 355, 360 and 375~MHz. The grey scale is
         polarized intensity $P$, oversampled 5 times, where white is
         high intensity. Maximum $P$\/ (in the 375~MHz band) is
         approximately 13~K. Superimposed are polarization
         pseudo-vectors, where each line denotes an independent
         synthesized beam.}
\label{f4:angle}
\end{figure*}

\subsection{Polarization angle}

In Fig.~\ref{f4:angle} we show, for a 3\dg$\times$2\dg\ subfield
slightly northwest of the center of the mosaic, maps of polarized
intensity $P$ (grey scale) and polarization angle $\phi$ (superimposed
pseudo-vectors) in the 5 frequency bands. In the grey scale, which has
5 samples per linear beam width, white corresponds to high $P$.
The length of the vectors (one per beam) scales with polarized
intensity.  The amount of structure in polarization angle is highly
variable. At high polarized intensities, the polarization angles
mostly vary quite smoothly. However, there are also abrupt changes on
the scale of a beam, the most conspicuous of which give rise to the
depolarization canals.

\begin{figure*}
  \centering
  \hbox{\psfig{figure=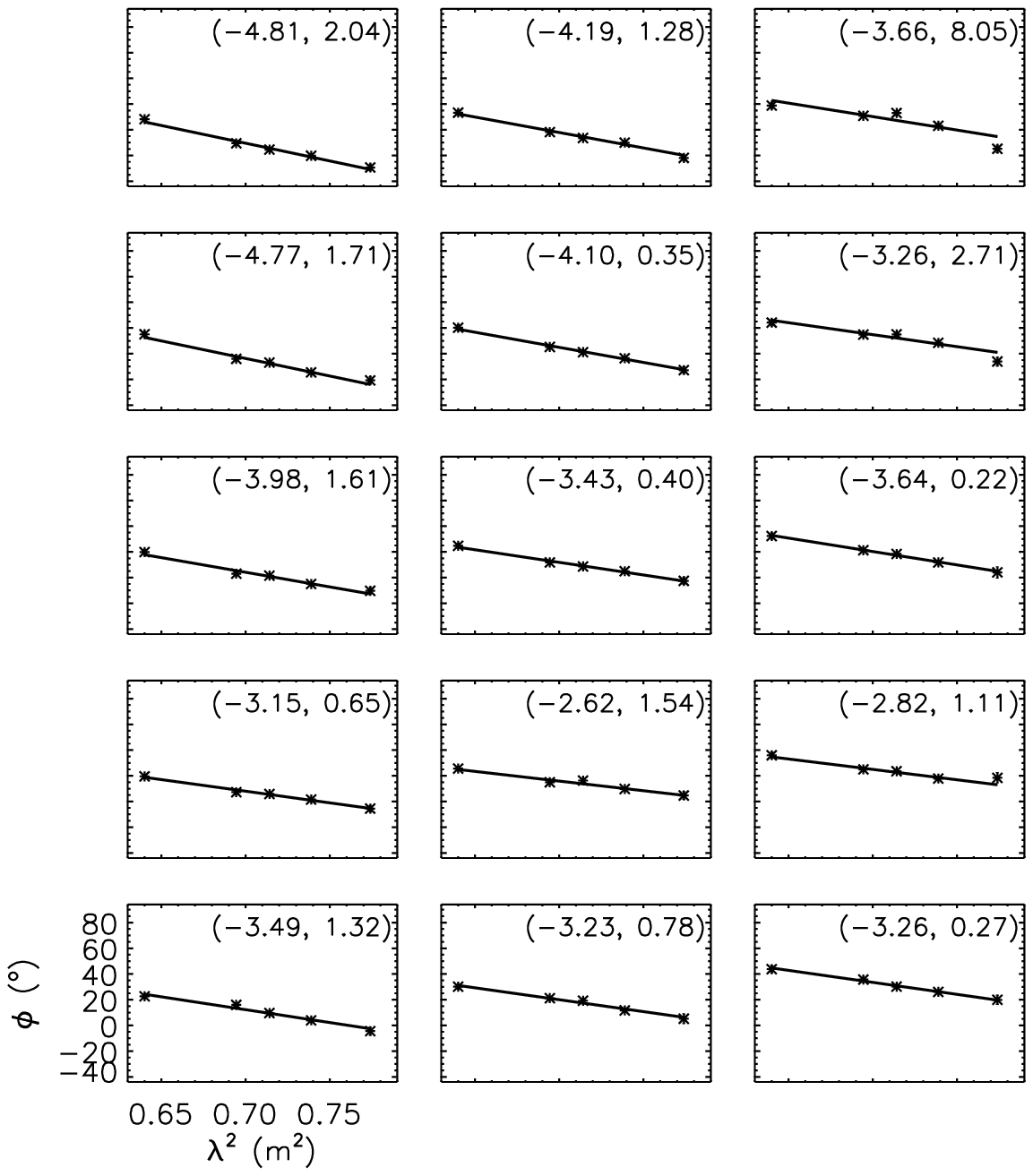,width=0.5\textwidth}
        \psfig{figure=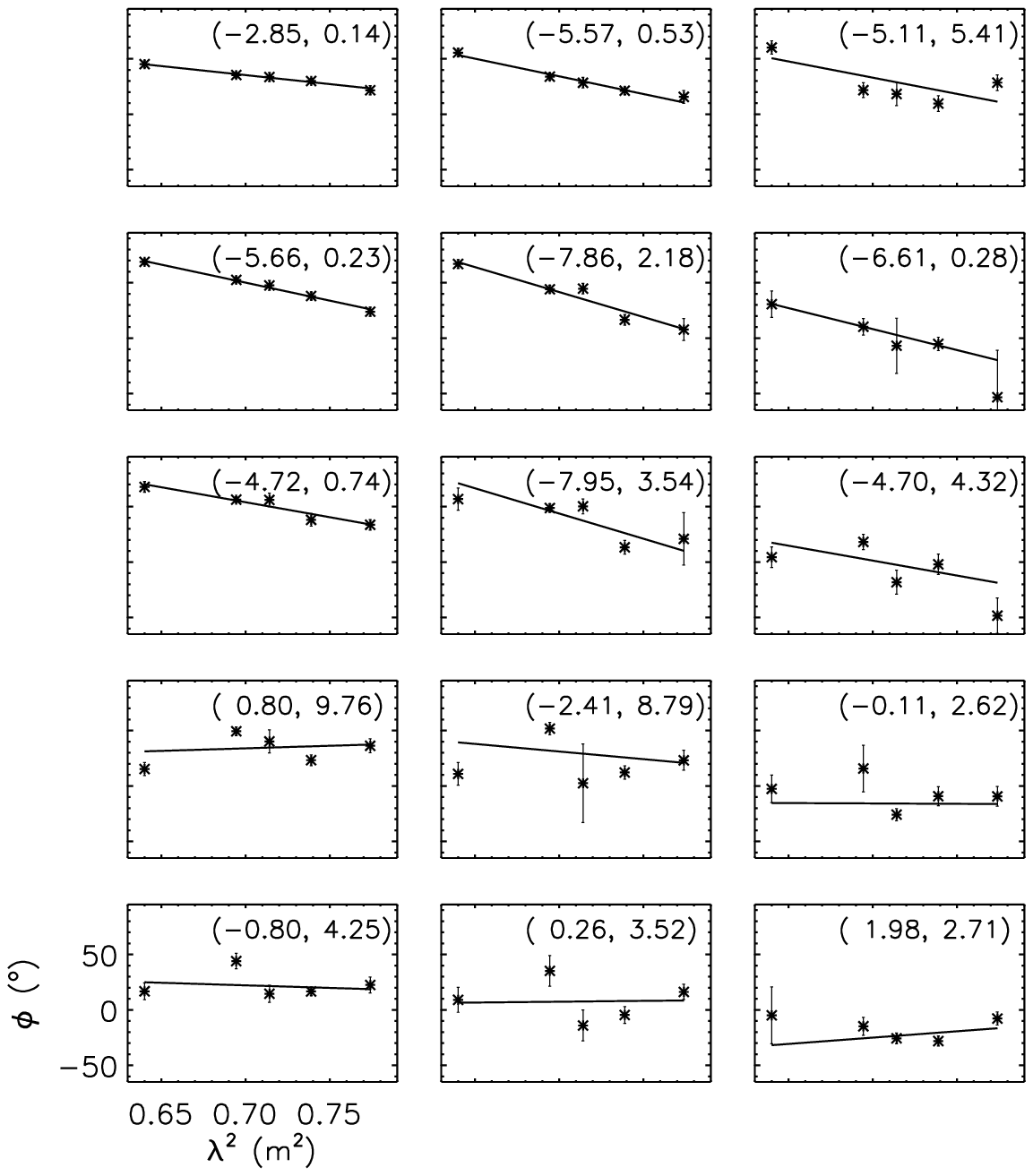,width=0.5\textwidth}}
  \caption{Two typical parts of 3$\times$5 beams in the observed
           field, represented in small graphs of polarization angle
           $\phi$ against $\lambda^2$, one graph per independent
           synthesized beam.  The left 3$\times$5 graphs are taken
           from a part of the field with high polarized intensity
           ($\sigma \ga 15$ per frequency band), the right part from a
           region of lower intensity ($\sigma \approx$~5). The values
           within parentheses above each graph are $RM$ and reduced
           $\chi^2$, respectively. In general, $RM$s are more regular
           and alike over several beams in high polarized intensity
           regions, but there is coherence in $RM$ in lower $P$
           regions as well.}
\label{f4:pl2}
\end{figure*}

\section{Faraday rotation}
\label{s4:rm}

\subsection{The derivation of rotation measures}

The rotation measure ($RM$) of the Faraday-rotating material follows
from $\phi(\lambda^2) \propto RM\,\lambda^2$.  As $\phi = 0.5\,\arctan
(U/Q)$ \footnote{See footnote in Sect.~\ref{s4:obs}.}, the values of
polarization angle are ambiguous over $\pm n\, 180^{\circ}$ ($n =
1,2,3,\ldots$) because a value of angle($Q,U$) is ambiguous
over $\pm n\, 360^{\circ}$. For most beams, it was possible to
obtain a linear $\phi(\lambda^2)$-relation by using angles with a
minimum angle difference between neighboring frequency bands.  Only in
about 1\% of the data, addition or subtraction 180\dg\ to the
polarization angle in at least one of the frequency bands was needed.
One could increase $RM$s by adding or subtracting multiples
of 180\dg, but that would lead to $RM$s in excess of 100~\radm.  If
such high $RM$s were real, bandwidth depolarization (see
Sect.~\ref{s4:depol}) would have totally annulled any polarized
signal. In addition, the $RM$s we obtain using the criterion of minimum
angle differences (of order of 10 \radm) were also obtained in other
observations in this direction (Bingham \& Shakeshaft 1967, Spoelstra
1984). So we conclude there are hardly any 180\dg-ambiguities in
polarization angle and we use the minimum angle difference between
frequency bands to determine $RM$s.

In Fig.~\ref{f4:pl2} we show $\phi(\lambda^2)$-plots for small arrays
of contiguous, independent beams in typical subfields of the
5\arcmin-resolution map. In parentheses are the values of the $RM$ and
the reduced $\chi^2$ of the linear $\phi(\lambda^2)$-fit.  The
leftmost 3$\times$5 plots are from a region with high $P$ ($\sigma
\ga$ 15 in all frequency bands), while the rightmost plots are from a
region with lower $P$ ($\sigma \approx 5$).  In general, the
$\phi(\lambda^2)$-relation is closer to linear in high $P$\/ regions
than in lower $P$\/ regions, but in the larger part of the Auriga
field $RM$s are very well-determined.

\subsection{The distribution of rotation measures}

\begin{figure*}
   \begin{center}
      \hbox{\psfig{figure=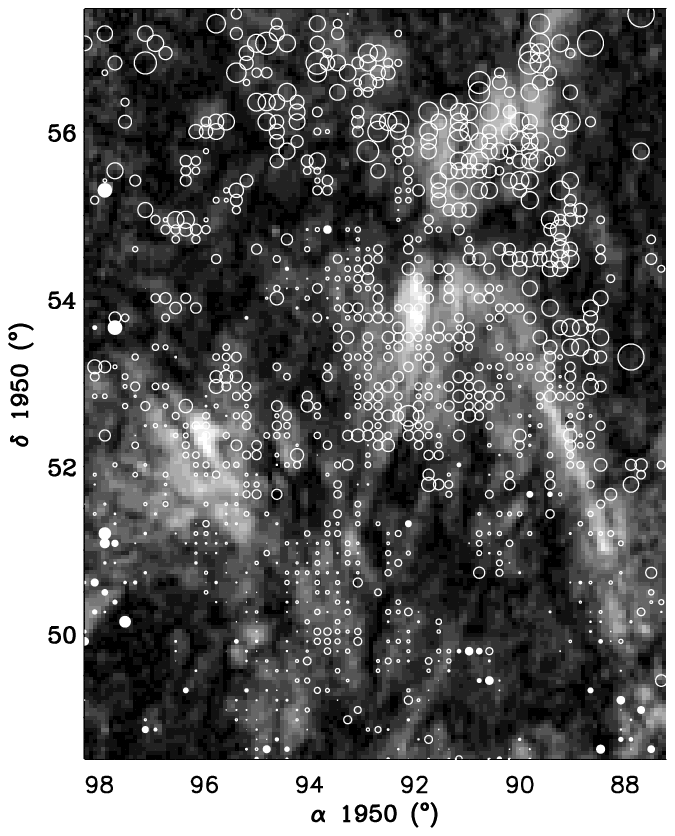,width=0.5\textwidth}
            \psfig{figure=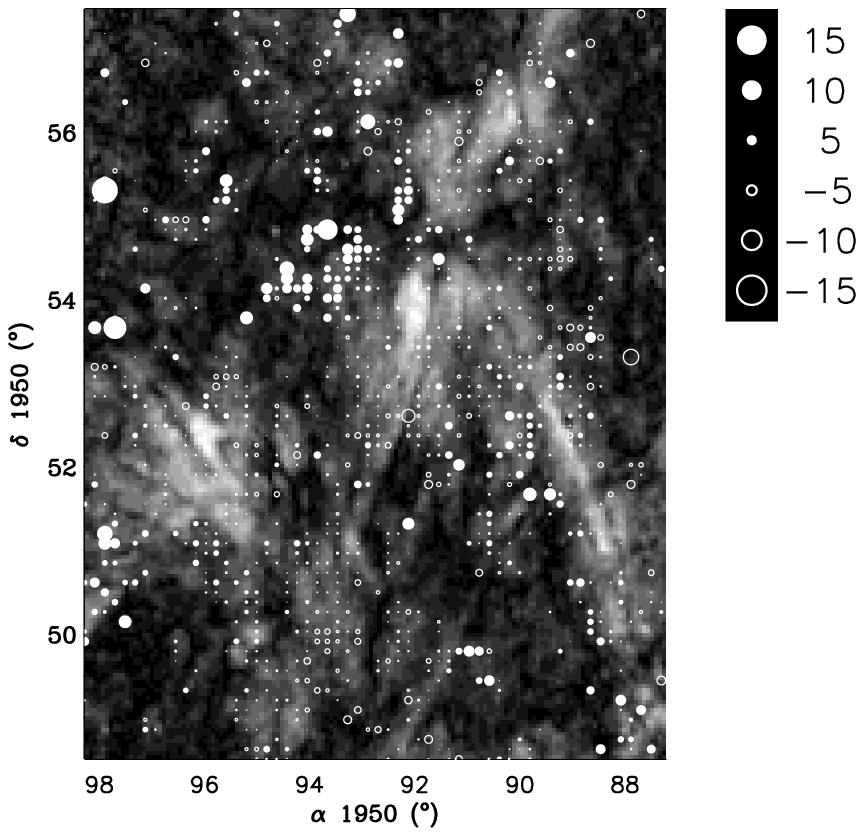,width=0.5\textwidth}}
      \caption[]{Rotation measures given as circles with a diameter
     proportional to the magnitude of the $RM$, superimposed over
     $P$\/ at 349~MHz in grey scale. The scaling is in \radm. $RM$s
     are only shown if $P > 5\sigma$ and reduced $\chi^2 < 2$, and in
     both directions, only every second independent beam is plotted
     for clarity. The right map shows the same data where the
     best-fitting linear gradient in $RM$ is subtracted.}
   \label{f4:rmmap} 
   \end{center}
\end{figure*}

The $RM$s in the Auriga-field are plotted as circles in
Fig.~\ref{f4:rmmap}. Filled circles denote positive values, open
circles negative, and the diameters of the circles are proportional to
$RM$. For clarity, we only show the $RM$s for one in four independent
beams. In addition, $RM$s are only shown when they were reliably
determined, i.e.\ for the average polarized intensity $\left<P\right>
> 5\sigma$, and for reduced $\chi^2$ of the fit $< 2$.

The left panel shows $RM$ as determined from the observations.  Most
$RM$s are negative; the few positive values occur for $\delta \la
52$\dg, while the largest negative values occur for $\delta \ga
55$\dg. We modeled this systematic variation by a large-scale linear
gradient, and find a gradient of about 1~\radm\ per degree, with the
steepest slope along position angle (from north through east) of
$-20\dg$.

\begin{figure}
\centering
\psfig{figure=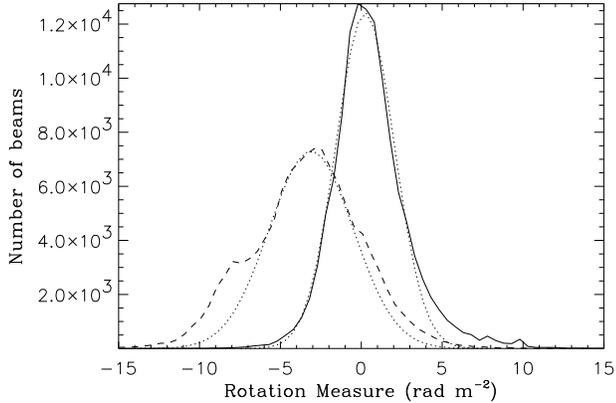,width=.45\textwidth}
\caption{Histogram of $RM$s in the Auriga field. The dashed line gives
           the distribution of all ``reliably determined'' $RM$s,
           i.e. where $P > 5\sigma$ and reduced $\chi^2 < 2$. The
           solid line gives the reliably determined $RM$ distribution
           after subtracting a best-fit linear $RM$ gradient from the
           data. Dotted lines show Gaussian fits to the two
           histograms.}
\label{f4:rmhist}
\end{figure}

The average $RM$, i.e.\ the $RM$ value in the center of the fitted
$RM$-plane, over the field is $RM_0 \approx -3.4$~\radm.  Subtracting
the best-fit gradient from the $RM$ distribution yields the $RM$s
shown in the right-hand panel in Fig.~\ref{f4:rmmap} with only
small-sale structure in $RM$.  The distribution of small-scale $RM$s
(i.e.\ $RM$ where the large-scale gradient is subtracted) is
significantly narrower than the total $RM$ distribution (\srm = 1.4~K
instead of 2.3~K), and is more symmetrical, as is shown in
Fig.~\ref{f4:rmhist}, where the two histograms are compared.

The decomposition of the $RM$-distribution into a constant component, a
gradient, and a small-scale component is not physical, as there is
probably structure on all scales.  However, for our purpose it is a
good approximation, and it allows us to estimate the large-scale and
random components of the Galactic magnetic field, see
Sect.~\ref{s4:disc}.

\subsection{Structure in  $RM$  on arcminute scales}

\begin{figure}
\centering
\hspace{-0.1cm}
   \psfig{figure=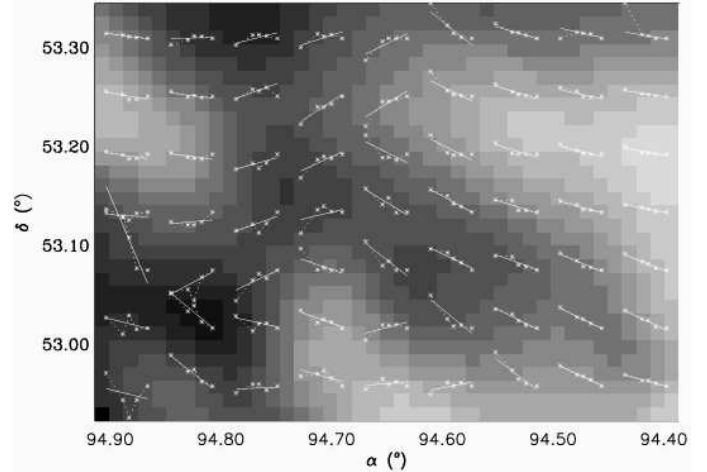,width=.5\textwidth} 
\caption[]{White symbols and lines are plots of $\phi$ against
   $\lambda^2$ and their linear fits, one for each independent beam,
   so that the slope is $RM$. Sudden $RM$ changes occur over one beam
   width. The grey scale is polarized intensity at 349 MHz oversampled
   by a factor of 5. Maximum $P$\/ (lightest grey) is $\sim$~40
   mJy/beam, while in the bottom left pixel $P$\/ is set to zero for
   comparison.}
\label{f4:RRfield}
\end{figure}

Although the $RM$-distribution shows structure on many scales, the
most intriguing changes occur on the scale of the beam, as illustrated
in Fig.~\ref{f4:RRfield}. In the figure, we display a 9$\times$7 array
of $\phi(\lambda^2)$-plots, overlaid on a grey scale representation of
$P$ at 349~MHz. Although $RM$ varies quite smoothly over most of the
area, there are also abrupt changes, which frequently involve a change
of the sign of $RM$ (e.g.  at $(\alpha, \delta) =$ (94.60, 53.00)\dg\
or (94.68, 53.20)\dg).  Abrupt $RM$ changes over one beam with the
right magnitude to cause an angle change of $\pm (n+1/2)\;180$\dg\
cause depolarization canals. As $RM$ is an integral over the entire
line of sight, it is difficult to understand such changes in $RM$ as
resulting from structure in the magnetic field and/or distribution of
electrons. As a matter of fact, numerical simulations of the warm ISM
show that beam depolarization can enhance $RM$ differences and greatly
steepen the gradient (Haverkorn 2002).  In addition, models of a
medium that produces Faraday rotation and that emits synchrotron
radiation show that the $RM$ can change sign without a corresponding
change in the direction of the magnetic field (Sokoloff et al.\ 1998,
Chi et al.\ 1997).

\section{The structure in $P$, and the implied properties of the ISM}
\label{s4:depol}

As $I$\/ and $P$\/ are uncorrelated, the structure in $P$\/ cannot be
caused by structure in emission, but instead is created by several
instrumental and/or depolarization effects. We discuss these processes
in detail in two forthcoming papers for several fields of observation,
among which the Auriga field (Haverkorn et al.\ 2003b,c). Here we
summarize them, and discuss the results for the Auriga region.

Structure in $P$ could be induced by the insensitivity of the WSRT to
large-scale structure, as a result of missing short-spacing
information. However, it is very unlikely that this effect is
important in the Auriga observations, because the distribution of
$RM$s per pointing center is too wide to allow offsets.  If the
distribution of $RM$s in a Faraday screen is too wide, the
distributions of $Q$ and $U$ of a polarized background that is
Faraday-rotated by the screen have averages very close to zero (even
if the background was uniformly polarized). As a result, the missing
short-spacing information does not produce offsets. For our
frequencies the latter is true if $\sigma_{RM} \ga 1.8$~\radm. In the
Auriga region, this condition is met in essentially all of the
subfields of the mosaic. Furthermore, a test in which offsets are
added (described in the papers referred to above) to produce better
linearity of the $\phi(\lambda^2)$-relation, did not result in
reliable non-zero offsets.

Structure in $P$ could also be generated by non-constant, but high,
$RM$ values as a result of bandwidth depolarization.  Summation of
polarization vectors with different angles in a frequency band will
reduce the polarized intensity. The amount of depolarization depends
on the spread in polarization angle within the band which, in turn,
depends on the $RM$ and the width of the frequency band. For the $RM$s
of order 10~\radm\ and our bandwidth of 5~MHz the effect is
negligible.

A special kind of structure in $P$ is due to beam depolarization,
viz.\ the cancellation of polarization by vector summation across the
beam defined by the instrument. This is a 2-dimensional summation of
polarization vectors across the plane of the sky, which is probably
responsible for the depolarization canals. It certainly cannot 
produce all of the structure in $P$ because it occurs on a single,
small scale, whereas $P$ has structure on a large range of scales.

Finally, depth depolarization occurs along the line of sight in a
medium that both emits synchrotron radiation and produces Faraday
rotation. It must be responsible for a large part of the structure in
$P$. The ISM in the Galactic disk most likely is such a medium. The
Galactic synchrotron emission has been modeled by Beuermann et
al. (1985) as the sum of two contributions: the thin disk with a half
equivalent width of $\sim$~180~pc, which coincides with the stellar
disk, and a thick disk of half equivalent width $\sim$~1800~pc (where
the galactocentric radius of the sun is assumed $R_{\odot} = 10$~kpc).
Thermal electrons exist also out to a scale height of approximately a
kpc, contained in the Reynolds layer (Reynolds 1989, 1991). So up to a
height of about a kpc, the medium probably consists of both emitting
and Faraday-rotating material.  In such a medium, the polarization
plane of the radiation emitted at different depths is Faraday-rotated
by different amounts along the line of sight. Therefore the
polarization vectors cancel partly, causing depolarization along the
line of sight even if the magnetic field and thermal and relativistic
electron densities are constant (differential Faraday rotation, see
Gardner \& Whiteoak 1966, Burn 1966, Sokoloff et al.\ 1998). If the
medium contains a random magnetic field, the polarization plane of the
synchrotron emission will vary along the line of sight causing extra
depolarization in addition to the differential Faraday rotation
(called internal Faraday dispersion).  We will refer to the
combination of the two mechanisms along one line of sight as depth
depolarization, thus adopting the terminology proposed by Landecker et
al. (2001).

Depth depolarization is the dominant depolarization process in the
Auriga field. We constructed a numerical model in Haverkorn et al.\
(2003c) to evaluate if depth depolarization can indeed be held
responsible for most of the structure in $P$, in the Auriga region as
well as in the other region that we studied. At the same time, the
model makes it possible to estimate the conditions in the warm
component of the ISM. The model consists of a layer corresponding to
the thin Galactic disk, where random magnetic fields on small scales
are present, and a constant background polarized radiation.

In the model, the thin disk (with height $h$) is divided in cells of a
single size $d$, which contain a random magnetic field component
$B_{ran}$ and a regular magnetic field $B_{reg}$.  The amplitude of
$B_{ran}$ is constant, but in each cell its
direction is randomly chosen. Each cell emits synchrotron radiation,
proportional to $B_{\perp}^2 =(B_{ran,\perp})^2 + (B_{reg,\perp})^2$,
where we add energy densities because the two components are
physically separated different fields. Only a fraction $f$ of the
cells contains thermal electrons, and therefore Faraday-rotates all
incoming radiation (to represent the filling factor $f$ of the warm
ISM).

We distinguish several kinds of model parameters. First, we have input
parameters with values taken from the literature; these are thermal
electron density $n_e = 0.08$~cm$^{-3}$, filling factor $f = 0.2$
(both from Reynolds 1991), and height of the Galactic thin disk $h =
180$~pc (Beuermann et al.\ 1985). Secondly, there are constraints from
the observations of the Auriga field, viz.\ the width, shape and mean
of the $Q$, $U$, $I$\/ and $RM$ distributions. In addition, we define
parameters without observational constraints, such as cell size $d$
and emissivity per cell, which we vary in a attempt to constrain them
from the model given the observational constraints.  Finally, there
are parameters that can be adjusted and optimized for each allowed
cell size and emissivity per cell, like the random and regular
magnetic field components and background intensity.

Using the observations of the Auriga region to constrain the
parameters in the depth depolarization model, we obtain the following
estimates. The cell size $d$ is probably in the range from 5 - 20~pc,
the random component of the magnetic field $B_{ran}$ is about 1~$\mu$G
for a cell size of 5~pc or larger, and up to 4~$\mu$G for smaller cell
sizes. With the estimate of the large-scale, regular field
($\sim$~3~$\mu$G) this leads to a value of about 0.3 to 1.3 for
$B_{ran}/B_{reg}$. This is lower than most other estimates of
$B_{ran}/B_{reg}$ in the literature, which seem to indicate a value
larger than 1. One possible explanation for this discrepancy is the
fact that the Auriga region is not very large. Therefore random
magnetic field components that project to angular scales larger than a
few degrees will have been interpreted as components of the regular
field. If such large-scale random components exist, this effect will
artificially decrease $B_{ran}/B_{reg}$. Secondly, the Auriga field
was selected for its conspicuous polarization structure, and high
polarized intensity in general indicates a more regular magnetic field
structure. Finally, the Auriga region is observed in an inter-arm
region. The next spiral arm, the Perseus arm, is more than 2~kpc away,
and we expect that at these frequencies, all polarized emission from
the Perseus arm is depolarized. There are indications that in the
inter-arm region, the regular magnetic field component is higher than the
random magnetic field component (Han \& Qiao 1994, Indrani and
Deshpande 1998, Beck 2001).

A cell of size 10~pc located at the far edge of the thin disk still
subtends a little less than a degree in our maps. We see structure on
smaller scales, so that cells of approximately a parsec have to be
present as well. Indeed, there are many indications that structure
exists in the ISM on many scales (Armstrong et al.\ 1995, Minter \&
Spangler 1996).

\begin{figure}
  \centering
  \psfig{figure=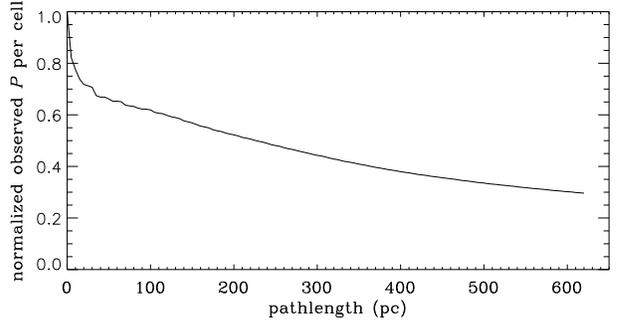, width=0.45\textwidth}
      \caption[]{The amount of polarization contributed by a certain
      cell, against the distance to that cell, in a single-cell-size
      model of polarized radiation propagating through a
      magneto-ionized medium.}
  \label{f4:polhor}
\end{figure}

The distance out to which polarized radiation is not depolarized by
depth depolarization, as derived from the model, is shown in
Fig.~\ref{f4:polhor}.
The total path length through the medium containing small-scale
structure was about 600~pc, which corresponds to a height of the thin
disk of about 150~pc for the Auriga latitude of $\sim$~16\dg.
Radiation emitted at large distances is depolarized more than emission
from nearby, but there is no definite polarization horizon due to
depth depolarization. Beam depolarization can cause a polarization
horizon, as structure further away will be on smaller angular scales,
which causes more beam depolarization (Landecker et al.\ 2001).

\section{Polarized extragalactic point sources}
\label{s4:sources}

\begin{table}
   \caption[]{Extragalactic sources in the Auriga region. The second
   column gives position, and the third $RM$s. Reduced $\chi^2$ of the
   $\phi(\lambda^2)$ relation is given in column 4. Columns 5, 6 and 7
   give resp. $P$, $I$\/ (both in mJy/beam) and percentage of polarization
   $p$ averaged over the five frequency bands.}
   \label{t4:src}
   \[
   \begin{array}{rrrrrrr} \hline
   \noalign{\smallskip} \mbox{No.} & (\alpha,\delta)\;(\;\dg,\;\dg)&
   \multicolumn{1}{c}{RM\; \mbox{(rad~m$^{-2}$)}} &
   \multicolumn{1}{c}{\;\;\;\chi^2} &
   \multicolumn{1}{c}{\;\;\left<P\right>\;} &
   \multicolumn{1}{c}{\;\;\left<I\right>\;} &
   \multicolumn{1}{c}{\;\;\left<p\right>} \\
   \noalign{\smallskip}
   \hline
   \noalign{\smallskip}
1\hspace{0.3cm} & \mbox{[ 97.9, 55.3 ]} &  7.4 \pm 0.2 &
\hspace{0.3cm}12.9 & 45.6 & 2021 & 2.3 \\
2\hspace{0.3cm} & \mbox{[ 97.7, 53.6 ]} & 12.9 \pm 0.3 & 
\hspace{0.3cm}3.1 & 14.9 &  216 & 6.9 \\ 
3\hspace{0.3cm} & \mbox{[ 96.6, 53.3 ]} &  6.6 \pm 0.4 & 
\hspace{0.3cm}1.7 & 15.0 &  224 & 6.7 \\ 
4\hspace{0.3cm} & \mbox{[ 96.5, 54.1 ]} &  5.7 \pm 0.4 &
\hspace{0.3cm} 3.7 & 11.9 &  380 & 3.1 \\
5\hspace{0.3cm} & \mbox{[ 95.3, 50.8 ]} &  0.6 \pm 0.4 &
\hspace{0.3cm} 1.6 & 14.7 &  165 & 8.8 \\
6\hspace{0.3cm} & \mbox{[ 95.2, 50.8 ]} &  3.1 \pm 0.4 &
\hspace{0.3cm} 0.2 & 14.7 &  197 & 7.5 \\
7\hspace{0.3cm} & \mbox{[ 94.7, 54.8 ]} &  7.9 \pm 0.3 &
\hspace{0.3cm} 3.0 & 18.8 &  282 & 6.7 \\
8\hspace{0.3cm} & \mbox{[ 93.4, 55.6 ]} &  6.3 \pm 0.2 &
\hspace{0.3cm} 1.1 & 21.3 &  180 & 11.8 \\
9\hspace{0.3cm} & \mbox{[ 92.3, 52.9 ]} &  2.7 \pm 0.1 &
\hspace{0.3cm} 6.1 & 35.8 & 1222 & 2.9 \\
10\hspace{0.3cm}& \mbox{[ 90.3, 50.4 ]} &-10.2 \pm 1.4 &
\hspace{0.3cm} 4.4 &  7.4 &  168 & 4.4 \\
11\hspace{0.3cm}& \mbox{[ 89.3, 55.0 ]} &  3.0 \pm 0.4 &
\hspace{0.3cm} 1.4 & 11.6 &  340 & 3.4 \\
12\hspace{0.3cm}& \mbox{[ 89.2, 52.5 ]} & -8.1 \pm 0.1 &
\hspace{0.3cm}12.7 & 22.1 &  477 & 4.6 \\
13\hspace{0.3cm}& \mbox{[ 89.0, 52.3 ]} &-13.6 \pm 0.4 &
\hspace{0.3cm} 5.8 & 14   &  237 & 5.9 \\
   \noalign{\smallskip}
   \hline
   \end{array}
   \]
\end{table}

Thirteen polarized extragalactic point sources in the Auriga-field
were selected based on three selection criteria: (1) signal-to-noise
$\sigma > 3$ in $P$ at all frequencies, (2) the degree of polarization
$p$ is higher than 1\% at all frequencies (because a lower
polarization degree can also be caused by instrumental polarization,
see Sect.~\ref{s4:obs}), and (3) the source does not lie too far at
the edge of the field away from any pointing center, to limit the
contribution from instrumental polarization.  The 13 sources are all
detected in the 1\arcmin\ resolution data, and most of them are
unresolved at that resolution. Their $RM$s vary from approximately
--13 to~13~\radm.

The relevant properties of the sources are given in
Table~\ref{t4:src}, where the second column gives the right ascension
and declination, column 3 the $RM$, and column 4 the reduced $\chi^2$
of the $\phi(\lambda^2)$-fits.  Columns 5, 6, and 7 give the values of
$P$, $I$\/ and percentage of polarization $p$ respectively, averaged
over all frequency bands.
\begin{figure}
  \centering
  \psfig{figure=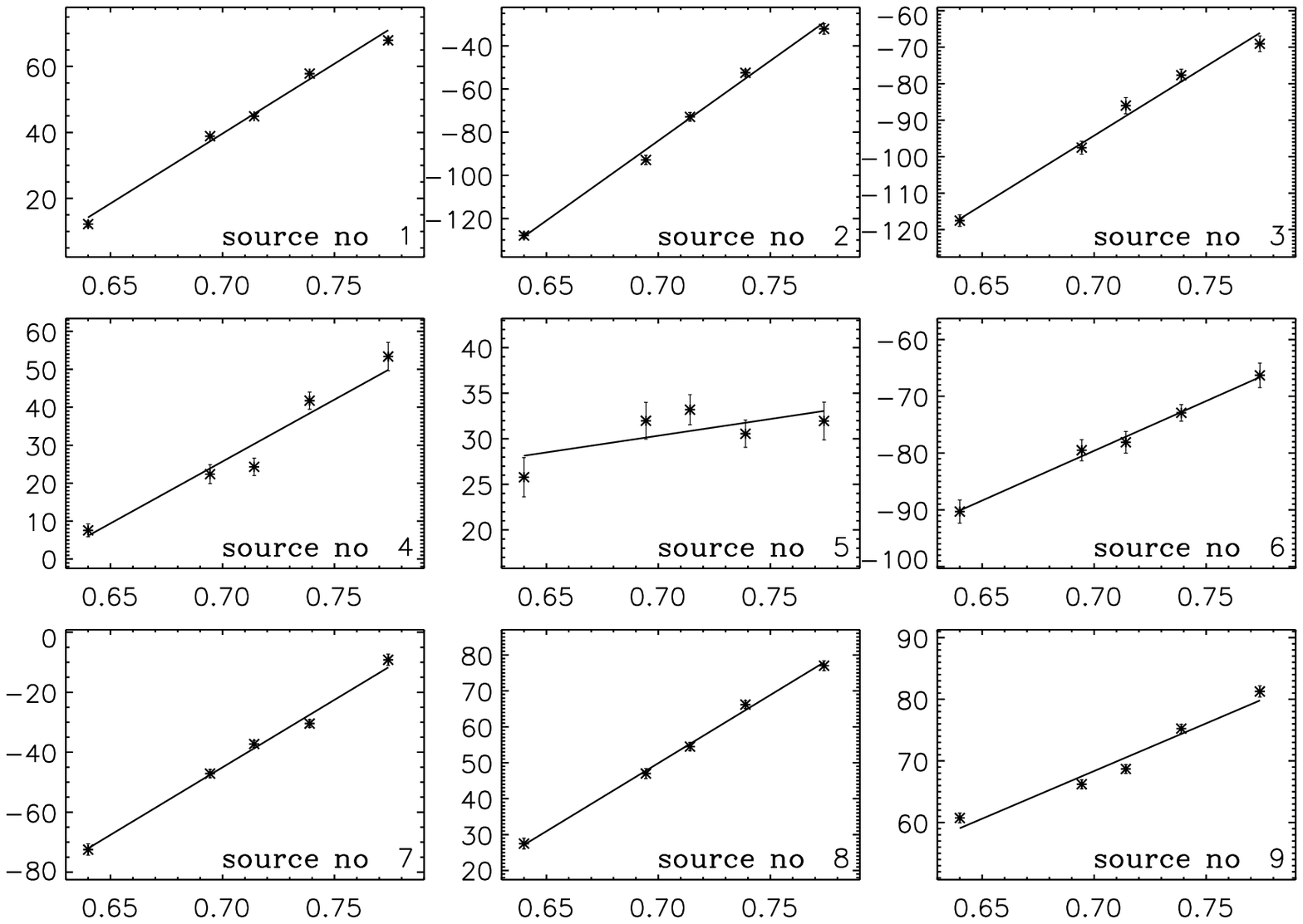,width=.5\textwidth}
  \psfig{figure=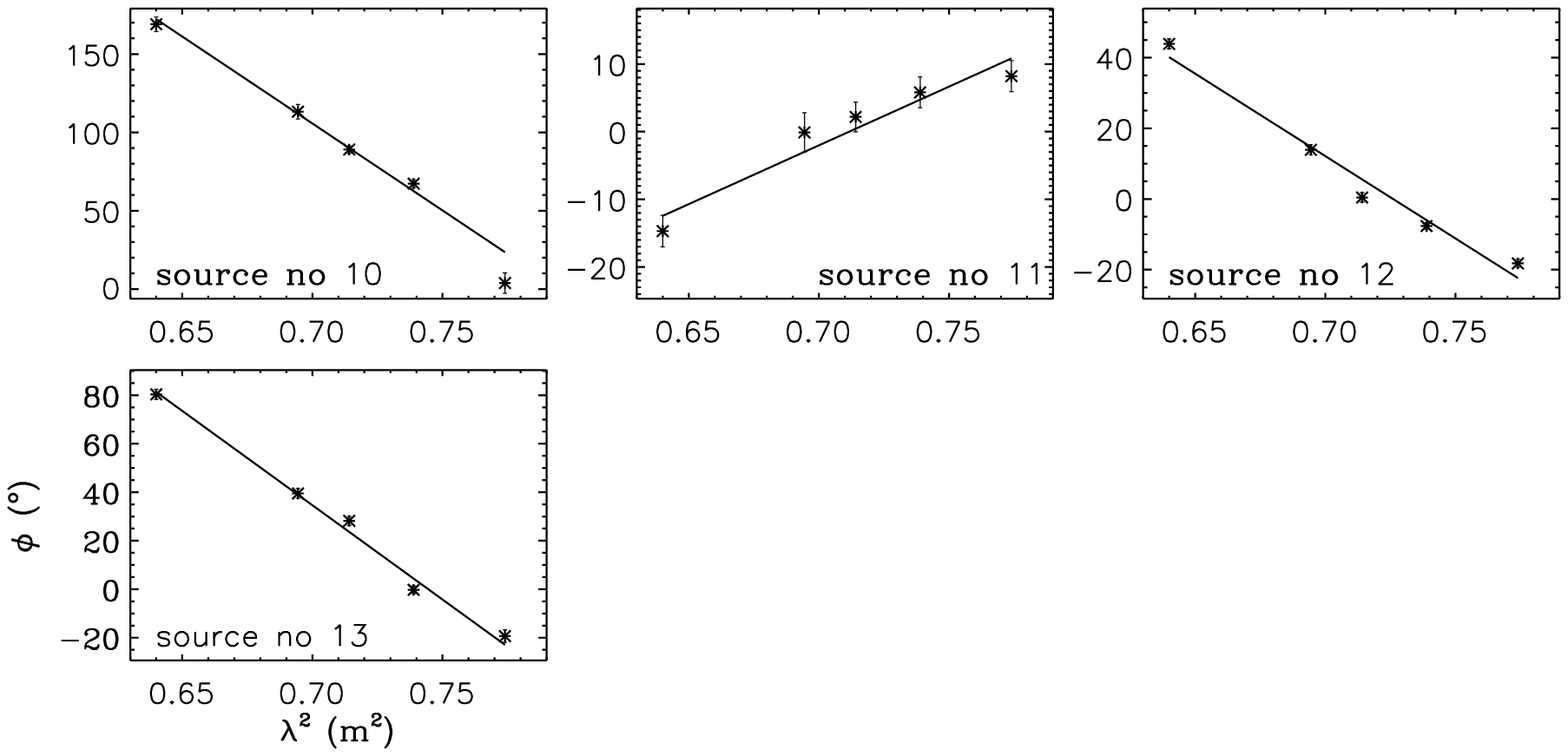,width=.5\textwidth}
  \caption[]{Plots of polarization angle $\phi$ against $\lambda^2$
             for the 13 polarized extragalactic sources in
             Table~\ref{t4:src}. Note that the scaling of the $\phi$-axis
             differs for each source.}
  \label{f4:src}
\end{figure}
Fig.~\ref{f4:src} shows the polarization angle $\phi$ plotted against
$\lambda^2$ for each point source. 
In Fig.~\ref{f4:circle}, we show the 13 sources
overlaid on a grey scale plot of polarized intensity at 349 MHz, with
their $RM$s indicated. The diameter of the symbol and its shading indicate
the value of $RM$. 

Sources 1 and 2 could have a significant contribution of instrumental
polarization, because they are observed in only one pointing center
and are located about 0.8\dg\ to 1\dg\ away from the pointing
center. Nevertheless, they show linear $\phi(\lambda^2)$-relations,
which indicates that instrumental polarization is not important. These
sources are therefore included in the analysis.

\begin{figure}
  \centering
  \psfig{figure=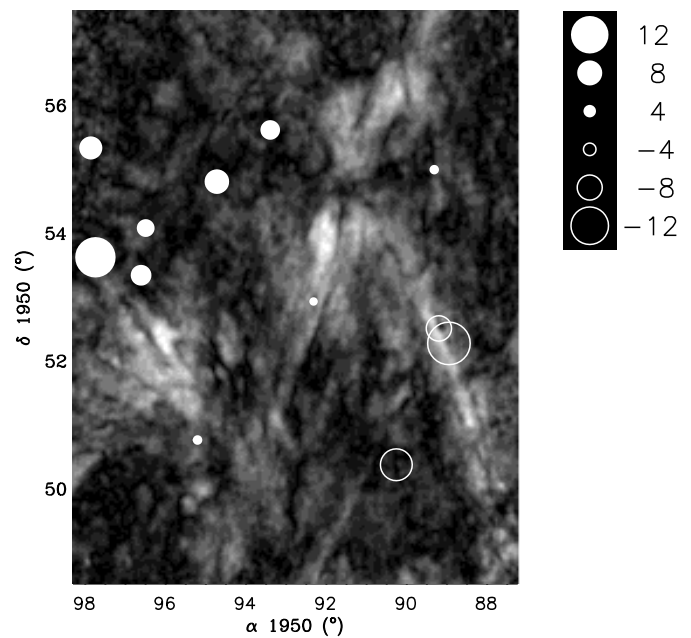,width=.5\textwidth}
  \caption{$RM$s of the 13 polarized extragalactic point sources in the
           Auriga field in circles, overlaid on $P$ at 349~MHz. Filled
           circles are positive $RM$s, open circles negative. The
           crosses denote the pointing centers, and the scaling is in
           \radm.}
  \label{f4:circle}
\end{figure}

The strong correlation of $RM$ across the sky indicates a Galactic
component to the $RM$s of the extragalactic point sources. The
best-fitting linear gradient to the $RM$s of the sources has a steepest
slope of 3.62~\radm\ per degree in position angle 72\dg, i.e. roughly
in the direction of increasing Galactic latitude (position angle
66\dg). The standard deviation of $RM$s around this gradient is
$\sigma_{RM} \approx 3.6$~\radm. This is
consistent with the result of Leahy (1987), who finds a typical
'internal' $RM$ contribution from the extragalactic source or a halo
around it of $\sim$~5~\radm.

\begin{figure}
  \centering
   \psfig{figure=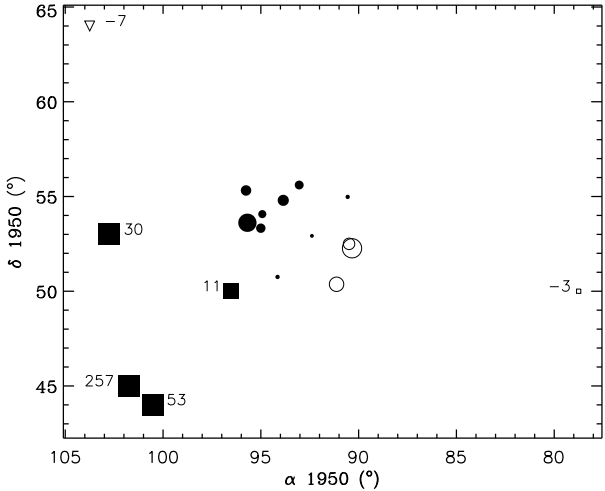,width=0.45\textwidth} 
   \caption[]{$RM$s of extragalactic point sources and of one
       pulsar. The circles are the point sources detected in the
       Auriga field, where the radius of the circle is proportional to
       the magnitude of $RM$.  Squares are extragalactic point sources
       detected by Simard-Normandin et al.\ (1981) and/or by Tabara
       \& Inoue (1980), and the one triangle in the field denotes the
       only pulsar near the Auriga field (Hamilton \& Lyne
       1987). Values of the $RM$ are written next to the sources, and
       for $RM$ $< 15$~\radm\ the radius of the symbol is proportional
       to $RM$. }
\label{f4:snk}
\end{figure}

The change of sign of $RM$ of the extragalactic sources within the
observed region indicates a reversal in the Galactic magnetic
field. (Unlike the diffuse emission, sign changes in $RM$s of
extragalactic sources cannot be due to depolarization effects.) The
reversal exists on scales larger than our field of view, as is shown
in Fig.~\ref{f4:snk}, where we combine the $RM$s of
our sources with those from the literature.  The circles are our
sources, the squares indicate sources that were detected by
Simard-Normandin et al.\ (1981) and/or by Tabara \& Inoue (1980), and
the one triangle indicates the only pulsar nearby (Hamilton \& Lyne
1987). The numbers next to the squares and triangle give the magnitude
of $RM$ of the source (the diameter of the symbol is only proportional to $RM$
up to $RM$~=~15~\radm). One source at $(\alpha,\delta) = (91.25\dg,
48\dg)$ was omitted, because Simard-Normandin et al.\ give $RM$~$=
34$~\radm\ and Tabara \& Inoue give $RM$~$=-64.9$~\radm.
The source distribution shows a clear magnetic field reversal, which
is not on Galactic scale but on smaller scales, as is shown in Figs.~1
and~2 of Simard-Normandin \& Kronberg (1980), which show $RM$s of
extragalactic sources over the whole sky.
 
The gradient in $RM$ measured in the extragalactic point sources is
completely unrelated and almost perpendicular to the structure in $RM$
from the diffuse emission. This can be explained from the widely
different path lengths that extragalactic sources and diffuse emission
probe. Diffuse emission can only be observed out to a distance of a
few hundred pc to a kpc, as radiation from further away will be mostly
depolarized. On the contrary, extragalactic sources are
Faraday-rotated over the complete path length through the Milky Way of
many kiloparsecs long. Therefore, the $RM$ structure from diffuse
radiation and from extragalactic sources gives information about
different regions in the Galaxy.

\section{Discussion}
\label{s4:disc}

\subsection{High $P$-structures and alignment with the Galactic plane}

\begin{figure}
  \centering
  \psfig{figure=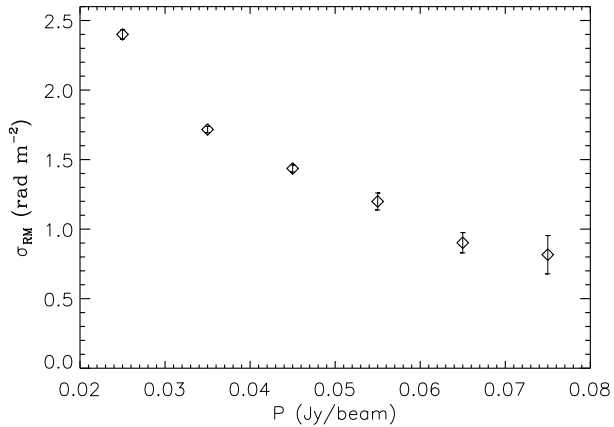,width=.45\textwidth}
  \caption{Width of a Gaussian fit to the $RM$ distribution \srm\ for a
           range of $P$, where $P$ is divided in bins of 0.01~Jy/beam.}
  \label{f4:srm_pi}
\end{figure}

\begin{figure}
  \centerline
  {\hbox{\psfig{figure=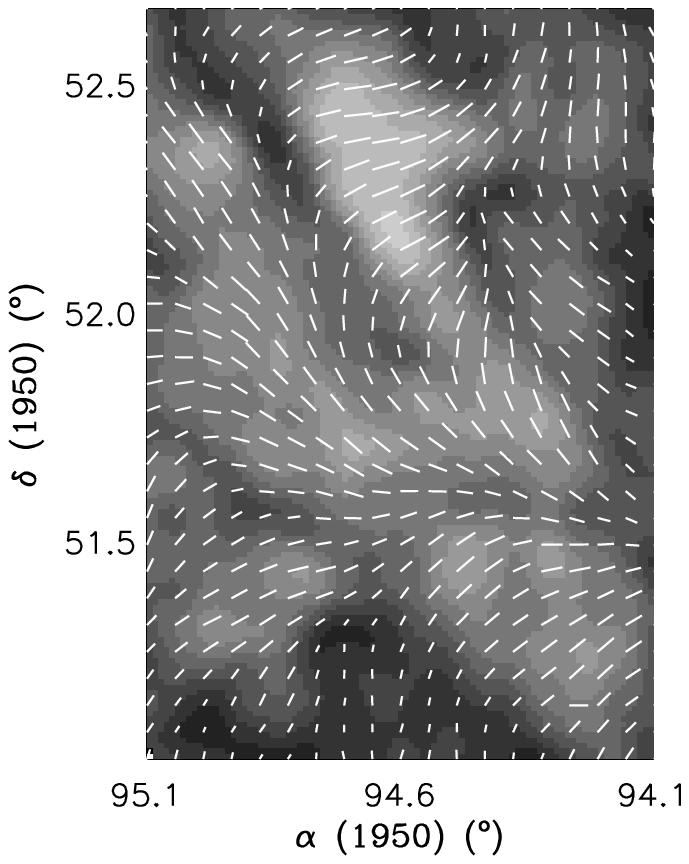,width=0.25\textwidth}
        \psfig{figure=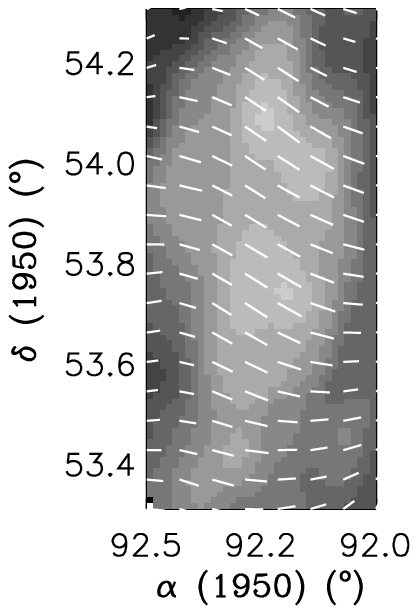,width=0.15\textwidth}}}
  \caption[]{Two areas in the $P$\/ map at 349~MHz with highest $P$
             ($> 80$~mJy/beam). Overlaid pseudo-vectors are
             polarization angles, showing a minimal change in angle
             for high~$P$.}
  \label{f4:angle_highp}
\end{figure}

The depth depolarization model described in Sect.~\ref{s4:depol}
only describes the properties of the ISM in a statistical manner, by
giving a representation of an average line of sight. Therefore, it
does not pretend to give a detailed description of the distribution of
$P$, as it is actually observed. Even the global properties of the $P$
distribution, like the general alignment with the Galactic plane, is
not part of the model.  This alignment is not only visible in the
Auriga field, but also in other regions observed with the WSRT
(Haverkorn et al.\ 2003a, Schnitzeler et al.\ in prep).  
The large-scale structures in
$P$ cannot be caused by substantial extra emission, because that
should be accompanied by corresponding structure in $I$, which is not
observed. Instead, where the polarized intensity is highest, the
depolarization is probably the least, so that there the ISM is
relatively transparent to polarized emission.  In general, high
depolarization is caused by a large amount of structure in $RM$, along
the line of sight and/or over the sky (the latter on the scale of the
beam or smaller).  This explanation of the structure of high $P$ is
borne out by the relation between polarized intensity $P$ and \srm,
shown in Fig.~\ref{f4:srm_pi}. This figure shows the width \srm\ of a
Gaussian fit to the $RM$ distribution in several intervals of
polarized intensity with a width $\Delta P = 0.01$~Jy/beam. Rotation
measure clearly varies more at lower polarized intensity, where only
well-determined $RM$s are used so that the effect cannot be due to
noise.  For the two intervals of highest $P$ (0.08~--~0.09~Jy/beam and
0.09~--~0.10~Jy/beam), not shown in Fig.~\ref{f4:srm_pi}, no Gaussian
fit could be made, as the distributions were bimodal.  The peaks in
these bimodal distributions correspond to two spatially coherent
structures, which are shown in the two panels of
Fig.~\ref{f4:angle_highp}.  In the region around $(\alpha,\delta) =
(94.6\dg,51.8\dg)$, $-1.5\la RM\la 0$~\radm, while in the region
around $(\alpha,\delta) = (92.2\dg,53.8\dg)$, $0\la RM\la
1.3$~\radm. Clearly, $RM$ is indeed very uniform in regions of high
polarization.  The low variation in $RM$ is also reflected in the
uniformity of the polarization angles.

A constant $RM$ over a certain area sets an upper limit on structure in
magnetic field and thermal electron density.  The fact that some of
the filamentary structure in $P$\/ is aligned with Galactic latitude
suggests a magnetic origin, although stratification of $n_e$ over
Galactic latitude can contribute as well. For variation of
polarization angle over the filament of less than about 20\dg\ (from
Fig.~\ref{f4:angle_highp}), $\Delta RM$ $\la 0.5$~\radm\ is needed.  If
magnetic field structure would be the same over the entire path length
towards the filament, and assuming $n_e = 0.08$~cm$^{-3}$, this
requires a magnetic field change $\Delta B_{\pl} \la 0.01~\mu$G with a
polarization depth of 600~pc. Even if the filamentary structures are
sheet-like with a constant magnetic field over a large part of the
path length, or if the polarization horizon is locally much shorter
than the highly uncertain value of 600~pc, the upper limit to $\Delta
B$ is very low. Therefore, this structure of high $P$\/ aligned with
Galactic latitude indicates the existence of long filaments or sheets
parallel to the Galactic plane of highly uniform magnetic field and
thermal electron density. As back- and/or foreground structure can
also imprint structure in $RM$, these regions may well be characterized
by low thermal electron density.

\subsection{The Galactic magnetic field}

The Auriga field of 7\dg$\times$9\dg\ represents only 0.1\% of the
sky, and is therefore by itself not very well suited for an analysis
of the global properties of the Galactic magnetic field. However, we can
estimate strength and direction of the regular component of the
Galactic magnetic field from the distribution of $RM$s of the diffuse
radiation and of the extragalactic sources over the field.

\begin{figure}
  \centering
  \psfig{figure=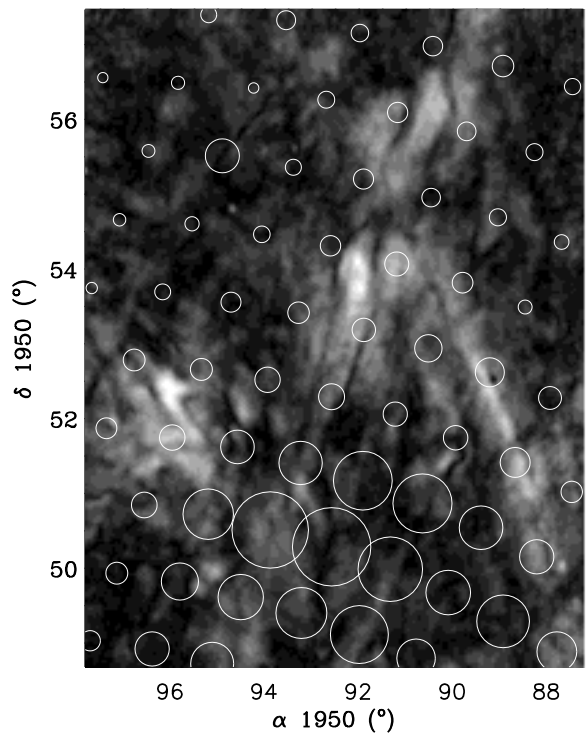,width=.45\textwidth}
  \caption{Superimposed circles are emission measures from the WHAM
           H$\alpha$ survey, to a maximum intensity of
           $\sim$~5~Rayleigh, at a resolution of one degree. The grey
           scale denotes $P$\/ at 349~MHz.}
  \label{f4:wham}
\end{figure}

First, we need to determine how much of the $RM$ gradient seen in the
diffuse emission is caused by $n_e$, and how much by
$B_{\pl}$. Independent measurements of $n_e$ can be obtained from the
emission measure $EM = \int n_e^2\;dl$, as measured with the Wisconsin
H$\alpha$ Mapper survey (WHAM, Haffner et al., in prep, Reynolds et
al.\ 1998), shown in Fig.~\ref{f4:wham}. As the resolution is about a
degree, the WHAM survey cannot be used to determine small-scale
structure in $n_e$, but gives an estimate of the global gradient in
$n_e$ over the field. From north to south, the H$\alpha$ observations
show an increase from about 1 to 5 Rayleighs (1 Rayleigh is equal to a
brightness of $10^6/4\pi$ photons cm$^{-2}$ s$^{-1}$ ster$^{-1}$, and
corresponds to an $EM$ of about 2 cm$^{-6}$ pc for gas with a
temperature T = 10000~K). The small enhancement in $EM$ at
$(\alpha,\delta) = (94.8\dg,55.5\dg)$ is probably related to the ancient
planetary nebula PuWe1 (PW1, Tweedy \& Kwitter 1996).  Note that the
H$\alpha$ emission probes a longer line of sight than the diffuse
polarization, and is velocity integrated over $-$80 to 80 km s$^{-1}$
LSR, so complete correspondence is not expected. However, a gradient
in $RM$ that was solely due to the $n_e$-gradient would have a sign
opposite to that of the observed $RM$-gradient. Therefore, the observed
gradient in $RM$ must be due to a gradient in the parallel component
of the magnetic field.  If the gradient in H$\alpha$ originates in the
same medium as the polarized radiation, the gradient in magnetic field
must even be large enough to compensate the gradient in electron
density in the opposite direction.

So the $RM$ gradient of the diffuse polarized emission of about 1~\radm\ per
degree in the direction of positive Galactic longitude must be caused
by a corresponding increase of the parallel component of the Galactic
magnetic field. The direction of the gradient in longitude and the
sign of the average $RM$ are consistent with a regular Galactic magnetic
field directed away from us in the second quadrant, as found from
other observations (see e.g.\ review by Han, 2001).  Attributing the
observed $RM$-gradient completely to the change in the parallel
component of the Galactic magnetic field, we can derive the strength
of its regular component.  The average $RM_0$ and the value of the $RM$
gradient give two independent estimates of the regular magnetic field,
if a certain pitch angle is assumed. Assuming a path length of 600~pc
and electron density $n_e = 0.08$~cm$^{-3}$, the two magnetic field
determinations agree on a regular magnetic field of $B_{reg} \approx
1~\mu$G for a pitch angle $p = -14$\dg, in agreement with earlier
estimates (Vall\'ee 1995).

The gradient in the $RM$s of the extragalactic sources does not show a
longitudinal component at all, contrary to that in the $RM$s of the
diffuse polarization observations. This indicates that other
large-scale magnetic fields become important, fields that vary on
scales larger than our field. Observations of halos of external
galaxies (e.g., Dumke et al.\ 1995) have shown that cell sizes in the
Galactic halo are much larger than in the disk, from 100
-~1000~pc. Even at a distance of 3000~pc, the size of the total Auriga
field is $\sim$~400~pc, so big cells in the halo could be responsible
for a considerable part of what we call the regular component of the
magnetic field, as measured in the $RM$s of the extragalactic sources.

We subtracted the gradient of the diffuse emission from the gradient
in the $RM$s of the extragalactic sources, after which a gradient in $RM$
of $\Delta RM$ $\approx 3.6$~\radm\ in position angle 54\dg\ remains
(only deviating by about 10\dg\ from the direction perpendicular to
the Galactic plane). This indicates a regular magnetic field of about
1~$\mu$G in this direction, assuming that $n_e$ and $B$ remain
constant over a total line of sight of 3000~pc. The magnetic field
strengths actually present in the medium will be higher if the field
varies along the line of sight.

\section{Conclusions}
\label{s4:conc}

Multi-frequency polarization observations of the diffuse Galactic
background yield information on structure in the Galactic warm gas and
magnetic field.

The multi-frequency WSRT observations of a region of size
$\sim$~7\dg$\times$9\dg\ at $l = 161$\dg, $b = 16$\dg\ in the
constellation Auriga show a smooth total intensity $I$, but abundant
structure in Stokes parameters $Q$\/ and $U$\/ on several scales, with
a maximum $T_{b,pol} \approx$~13~K (about 30\% polarization).
Filamentary structure up to many degrees long is present in $P$,
sometimes aligned with Galactic latitude. In addition, narrow,
one-beam wide depolarization canals are most likely created by beam
depolarization and may well indicate abrupt $RM$ changes. As the
polarization structure is uncorrelated to $I$, the structure in $P$\/
cannot be created by variations in synchrotron emission, but has to be
due to Faraday rotation and depolarization mechanisms.

Rotation measure maps show abundant structure on many scales,
including a linear gradient of $\sim$~1~\radm\ per degree in the
direction of positive Galactic longitude and an average $RM_0 \approx
-3.4$~\radm. The gradient is consistent with the regular Galactic
magnetic field if its strength is $B_{reg} \approx 1~\mu$G and the
field is azimuthally oriented with a pitch angle $p \approx
-14$\dg. Ubiquitous structure is present in the $RM$ map on beam size
scales (5\arcmin), indicating significant changes in magnetic field
and/or thermal electron density over very small spatial scales.

There are two dominant depolarization mechanisms which create
structure in $P$ in the Auriga field. First, beam depolarization
(i.e.\ depolarization due to averaging out polarization angle
structure within one beam width) most likely creates the canals, and
is important in regions of low $P$. Additional depolarization is
needed to explain the observed $P$ distribution, which is only
possible if the medium both Faraday-rotates and emits synchrotron
emission.  Then, emission originating in the medium can be depolarized
along its path towards the observer, so-called depth depolarization.
The polarization angle along the path can vary due to Faraday
rotation, or due to varying intrinsic polarization angle of the
emission. This indicates a varying magnetic field, which however is
constrained by the upper limit on structure in $I$.  A depth
depolarization model was constructed of a layer of cells with varying
magnetic fields and a constant background polarization. Constraints
from the Auriga observations give estimates for several parameters in
the ISM. The cell size of structure in the ISM is constrained to
$\sim$~15~pc, and the ratio of random to regular field is $0.7 \pm
0.5$.

Thirteen extragalactic sources in the Auriga field also show a
gradient in $RM$, but roughly in the direction of Galactic latitude,
which is perpendicular to the gradient in the diffuse emission. The $RM$
distributions from diffuse radiation and from extragalactic point
sources are so different because the diffuse radiation mostly probes
the first few hundred parsecs, whereas $RM$s from the point sources are
built up along the entire line of sight through the Milky Way. $RM$s of
extragalactic sources change sign over the field, which indicates a
local reversal of the magnetic field.

\section*{Acknowledgements}
We thank R. Beck, E. Berkhuijsen and F. Heitsch for critical reading
and useful comments.  The Westerbork Synthesis Radio Telescope is
operated by the Netherlands Foundation for Research in Astronomy
(ASTRON) with financial support from the Netherlands Organization for
Scientific Research (NWO). The Wisconsin H-Alpha Mapper is funded by
the National Science Foundation. MH is supported by NWO grant
614-21-006.

\end{document}